\documentclass[review,11pt]{elsarticle}

\usepackage{setspace}
\usepackage{amssymb}
\usepackage{amsmath}
\usepackage{amsthm}

\usepackage{booktabs}
\usepackage{graphicx}
\usepackage{multirow}
\usepackage{threeparttable}
\usepackage{tablefootnote}
\usepackage{float}

\newtheorem{assumption}{Assumption}
\newtheorem{theorem}{Theorem}

\usepackage{xcolor}
\usepackage[
colorlinks=true,
linkcolor=blue,
citecolor=blue,
urlcolor=blue
]{hyperref}
\usepackage{doi}
\biboptions{authoryear,round}

\usepackage{xr-hyper}
\journal{}

\begin{document}

\begin{frontmatter}
	
	
	
	\title{Large-Scale Asset Selection via Metric Dependence with Enriched High Frequency Information} 

	
\author[sustech]{Yangzhou Chen\fnref{equal}}
\author[hku]{Shuaida He\fnref{equal}}
\author[sustech]{Xin Chen\corref{cor1}}
\ead{chenx8@sustech.edu.cn}

\address[sustech]{Department of Statistics and Data Science, Southern University of Science and Technology, 1088 Xueyuan Avenue, Shenzhen, Guangdong 518055, China}
\address[hku]{School of Computing and Data Science, The University of Hong Kong, Pokfulam Road, Hong Kong 999077, China}

\fntext[equal]{These authors contributed equally to this work.}
\cortext[cor1]{Corresponding author.}

\begin{abstract}
Large-scale portfolio choice is highly sensitive to estimation error, making the preliminary asset selection essential in empirical implementation. Existing selection rules typically rely on scalar returns or low dimensional high frequency summaries, and thus discard intraday risk dynamics that may be relevant for risk adjusted allocation. 
We propose Metric Dependence Screening (MDS), an asset selection procedure that incorporates high frequency information as object valued data. Each asset day observation is represented as a point-curve object combining daily return with an intraday risk state curve, equipped with a weighted product metric that preserves both reward information and within day risk dynamics. MDS ranks assets by a Fr\'echet variation based dependence score, measuring how much a risk adjusted target explains the metric dispersion of the asset representations. This yields a simple two stage portfolio procedure: MDS first reduces the investable universe, and standard mean-variance or minimum variance allocation is then applied. We develop a target slicing estimator and establish concentration, sure selection, and rank consistency guarantees under $\alpha$-mixing time series dependence and ultrahigh dimensionality. 
Simulations show that MDS performs well across both Euclidean and non-Euclidean settings. Using high frequency data for $2938$ Chinese A-share stocks from July 2023 to December 2025, we demonstrate that MDS improves out of sample portfolio performance over return based and scalar dependence based benchmarks, highlighting the value of preserving intraday risk dynamics.

\end{abstract}
	
%

	
\end{frontmatter}



\section{Introduction}
\label{sec:Introduction}

Portfolio construction is a central problem in empirical finance because investors must allocate wealth across risky assets while balancing expected return and risk \citep{campbell2002strategic}. Mean-variance portfolio choice, developed by \citet{markowitz1952,markowitz2008portfolio}, provides a canonical framework that links optimal portfolio weights to the mean vector and covariance matrix of asset returns. Although the framework is conceptually simple, its empirical implementation depends heavily on the size of the opportunity set. In modern equity markets, the vast and evolving universe of tradable assets makes portfolio construction inherently high dimensional, placing dimensionality at the center of practical mean-variance implementations.
	
Implementing mean-variance portfolios requires estimating expected returns and return covariances and then using these estimates in optimization. In large asset universes, this task is inherently high dimensional because a large number of means and an even larger number of covariances must be estimated. As a result, portfolio weights can be highly sensitive to input estimation error. Early work studies the sampling behavior of mean-variance efficient weights and related inference problems \citep{jobson1980estimation,britten1999sampling}. \citet{best1991sensitivity}, \citet{chopra1993effect}, and \citet{michaud1989markowitz} show that errors in expected returns and covariances can have a large effect on mean-variance portfolios because optimization tends to amplify input noise. Related work also shows that extreme or poorly diversified efficient weights can arise in large cross sections \citep{green1992will}. \citet{kan2007optimal} show that accounting for parameter uncertainty can materially change the optimal portfolio and affect out of sample performance, while uncertainty averse formulations provide another route to more disciplined input use \citep{garlappi2007portfolio}. Empirical evidence further shows that simple allocation rules can be difficult to dominate out of sample when estimation risk is substantial \citep{demiguel2009optimal}. These concerns have motivated stabilization strategies such as Bayesian or shrinkage approaches to expected returns \citep{jorion1986bayes,frost1986empirical} and regularized estimation of covariance inputs \citep{ledoit2003honey,ledoit2003improved}.
	
Even with regularized mean-variance inputs, portfolio optimization over a very large and frequently rebalanced universe can still be impractical. A common solution is a two stage procedure. The first stage screens the universe down to a manageable subset. The second stage constructs portfolio weights on the selected assets through a constrained minimum variance or mean-variance program. The economic motivation is that high dimensional portfolio optimization is well known to be sensitive to estimation error, which can lead to unstable and extreme allocations. Portfolio constraints can mitigate this instability and improve out of sample performance by limiting the exposure of optimized weights to input noise \citep{jagannathan2003risk,fan2012vast}. This motivates treating the first stage formally as a high dimensional screening problem tailored to portfolio objectives. In this direction, \citet{wang2022asset} develops an important benchmark framework for asset screening based on a high frequency Sharpe ratio target and establishes finite sample sure screening and ranking consistency properties under realistic time series dependence. Their work provides a useful starting point for studying screening rules that are designed specifically for portfolio construction. Following \citet{wang2022asset}, we construct a risk adjusted target series $Y$ based on high frequency Sharpe information, so that the screening step is aligned with portfolio allocation rather than raw return prediction alone.
	
A key modeling choice in two stage portfolio construction is the information set used to represent each asset in the first stage ranking. In much of the existing empirical practice, the screening covariate is built from low dimensional, typically daily, summaries, and assets are ranked by marginal scores computed from daily return series, daily characteristics, or other scalar signals. This remains true in screening rules designed for dependent financial time series. For example, in \citet{wang2022asset}, the screening covariate $X$ is driven mainly by daily return information, while high frequency data enter primarily through a risk adjusted target constructed as a scalar summary. More broadly, even when portfolio objectives are explicitly adjusted for risk, intraday information is often incorporated through scalarization, for example through realized volatility, related jump robust variation measures, or one day risk adjusted scores \citep{andersen2001distribution,andersen2003modeling,barndorff2002econometric,barndorff2004power}. Such summaries have proved useful for volatility forecasting and risk managed allocation \citep{fleming2001economic,martens2007measuring,ait2008out}, but they can be restrictive in screening first pipelines. A large literature documents systematic intraday patterns, persistent diurnal structure, and substantial heterogeneity in high frequency volatility across assets and market conditions \citep{andersen1997intraday,bollerslev2000intraday}. When these within day shape features matter for subsequent allocation, reducing an intraday path to a few moments or a single realized measure may discard informative variation. As a result, the first stage ranking may be weakened, and the quality of the candidate set passed to the second stage optimizer may also decline.
	
The loss of intraday information is especially important in screening first portfolio construction, because the first stage determines the candidate universe passed to the optimizer. Formally, the first stage ranks assets through a marginal association score between an asset level representation and a target variable that summarizes the investor's objective. When the objective is adjusted for risk, representing assets only by close to close returns can create a mismatch in information. Assets with similar daily returns may still have very different intraday risk dynamics, and these differences may lead to materially different risk exposures over the holding period. This motivates enriching the screening information set so that it incorporates intraday risk states while remaining suitable for large universe ranking.
	
Motivated by this consideration, we enrich the screening covariate so that it reflects both reward and the intraday risk state without imposing an aggressive low dimensional reduction. 
Specifically, we represent asset $i$ on day $t$ by a point curve object
\[
X_{i,t}=\bigl(R_{i,t},\,v_{i,t}\bigr)\in \mathcal X,
\]
where $\mathcal X=\mathcal R\times\mathcal V$, $R_{i,t}$ is the daily return, and $v_{i,t}$ is the latent intraday spot volatility curve. Under the continuous time diffusion representation of the within day log price process, $v_{i,t}(u)$ is identified with the spot volatility $\sigma_{i,t}(u)$. The empirical implementation replaces this latent curve by a finite grid estimator constructed from high frequency returns. This construction preserves within day volatility shape information, including the timing and concentration of volatility bursts.
Such information is often lost when intraday observations are reduced to a single realized measure or a small set of handcrafted summaries. The representation $X_{i,t}=(R_{i,t},v_{i,t})$ therefore captures both the asset's daily reward and its intraday risk state. As a result, the first stage ranking is based on an information set that is more relevant for risk adjusted portfolio allocation.

The covariate $X_{i,t}$ is not a standard Euclidean vector. It is a point-curve object whose second component is functional, so it is more natural to treat it as a random object in a metric space. This view is consistent with a growing literature on the statistical analysis of non-Euclidean and object valued data, where inference is developed directly from the underlying metric structure rather than from a specific Euclidean embedding or a dimension reduction. Representative examples include Fr\'echet regression for random objects \citep{petersen2019frechet}, Fr\'echet sufficient dimension reduction for random objects \citep{ying2022frechet}, single index Fr\'echet regression \citep{bhattacharjee2023single}, and dimension reduction for Fr\'echet regression \citep{zhang2024dimension}.

Building on this metric space formulation, we analyze $X_{i,t}$ through the metric structure induced by the point-curve representation and propose a two stage metric dependence screening (MDS) procedure. In the first stage, we rank assets by a non-Euclidean correlation type dependence score defined on the metric space of point-curve objects and select a subset of suitable size. In the second stage, we construct portfolio weights on the selected subset using a mean-variance or minimum variance program, following the two stage portfolio construction framework of \citet{chen2017sure}. The dependence score is built from the Fr\'echet correlation coefficient of \citet{he2026frechetcorrelationcoefficientheterogeneous}. It is also related to a broader literature on dependence measures beyond linear correlation, including distance covariance and distance correlation \citep{szekely2007measuring,szekely2009brownian}, their extension to general metric spaces \citep{lyons2013distance}, kernel based measures such as HSIC \citep{gretton2005measuring,gretton2007kernel}, and ball correlation in Banach spaces \citep{pan2020ball}. The coefficient we adopt is defined through Fr\'echet variance in the underlying metric space and admits an explained variation interpretation. This makes the resulting ranking criterion economically transparent and close in spirit to Markowitz mean-variance reasoning: an asset receives a higher score when the risk adjusted target explains a larger reduction in the metric dispersion of its point-curve representation.

With this dependence score, the first stage produces a reduced universe on which standard second stage optimizers can be implemented more stably. Methodologically, this two stage design fits within the broader screening paradigm in high dimensional statistics, where marginal relevance ranking is used to reduce dimension before a more structured procedure is applied. Representative examples include sure independence screening and related developments for dependent data \citep{fan2008sure,fan2010selective,li2012feature}.


This paper proposes MDS for large-scale asset selection with enriched high frequency information, and its contributions are fivefold.
\begin{itemize}
    \item \textbf{Conceptually}, high frequency asset information is treated as object valued data rather than scalar summaries. 
    Each asset day observation is represented by a point curve object combining daily return and a latent intraday spot volatility curve. The associated weighted product metric balances the return coordinate and the intraday volatility curve coordinate, thereby preserving both reward information and within day risk dynamics. More broadly, many complex economic data objects may also be represented in similar non-Euclidean forms through appropriate metric structures.
    
    \item \textbf{Methodologically}, MDS provides a unified framework for asset ranking and selection based on a metric dependence score. The score is defined through Fr\'echet variation and measures how much the risk adjusted target reduces the dispersion of each asset representation. This metric perspective is in the same spirit as mean-variance analysis, while allowing the handling of object valued asset information through Fr\'echet variation. It also yields an interpretable ranking rule for a two stage pipeline in which MDS selects a reduced asset universe before standard portfolio allocation. Moreover, the procedure can be used beyond the specific point-curve metric adopted here.
    
     \item \textbf{Operationally},
     the metric dependence score is turned into an operational asset selection and allocation pipeline. In the first stage, MDS ranks and screens assets while allowing relevance to be assessed for nonvectorial covariates beyond marginal linear associations or coordinate specific Euclidean summaries. In the second stage, the selected universe is passed to standard mean-variance or minimum variance portfolio construction. This provides a practical bridge between object valued asset information and conventional portfolio allocation.

    \item \textbf{Theoretically}, built on the proposed weighted point-curve product metric, an asymptotic analysis is developed under $\alpha$-mixing time series dependence and ultrahigh dimensionality. Since the asset covariates are non-Euclidean objects without a standard vector structure, the MDS estimator is not a standard Euclidean marginal statistic. Its analysis requires delicately handling target slicing bias, empirical Fr\'echet mean estimation error, and dependent empirical fluctuations over the metric space. This unified control establishes concentration, sure screening, and rank consistency guarantees.

    \item \textbf{Empirically}, this paper provides a large-scale portfolio screening study using point-curve asset covariates to represent high frequency intraday risk information. Using high frequency data for $2938$ Chinese A-share stocks from July 2023 to December 2025, we show that MDS is feasible in a realistic ultrahigh dimensional asset universe and delivers clear out of sample portfolio gains relative to benchmark screening methods. These results highlight the practical value of preserving intraday risk information in asset selection.
\end{itemize}

The rest of this paper is organized as follows. Section~\ref{sec:methodology} introduces the point-curve covariate representation and the associated metric structure. It then develops the MDS coefficient, a feasible partition based estimator, and the resulting ranking rule. Section~\ref{sec:MDS-theory} establishes theoretical guarantees for MDS in dependent and ultrahigh dimensional settings, including concentration bounds, the sure screening property, and rank consistency. Section~\ref{sec:simulation} studies the finite sample performance of MDS through Monte Carlo experiments in both Euclidean and non-Euclidean settings. Section~\ref{sec:empirical} applies MDS to a large universe asset selection problem in the Chinese A-share market using high frequency data and examines the out of sample portfolio implications. Section~\ref{sec:Conclusion} concludes.
	
\section{Methodology}\label{sec:methodology}

This section introduces the main ingredients of the proposed screening procedure. We first define an asset level point-curve covariate together with its product metric and construct a one dimensional risk adjusted target series. We then develop the metric dependence score and the resulting MDS rule for asset ranking and selection.

\subsection{Setup: objects and target}\label{subsec:setup}
In the screening stage, each asset is represented by a covariate object $X_{i,t}$, and relevance is assessed relative to a one dimensional target series $Y_t$ that summarizes a risk adjusted investment objective. We first define the asset level point-curve covariate and the metric structure on its sample space. We then construct the target series from high frequency Sharpe based sorting. These objects serve as the basic ingredients for MDS developed in Section~\ref{subsec:mds}.
	
\subsubsection{Asset covariate $X$}\label{subsubsec:objects-X}

The screening stage uses an asset level covariate that combines a point valued return component with an intraday volatility state component. Let $\mathcal U\subset\mathbb R$ be a bounded interval denoting the intraday time domain for a trading day, equipped with a finite measure $\mu_{\mathcal U}$ such that $\mu_{\mathcal U}(\mathcal U)<\infty$. For each stock $i$ and trading day $t$, let $R_{i,t}$ denote the daily return.

To define the intraday component, we view the within day log price process as a continuous time diffusion. Specifically, on the intraday time domain $\mathcal U$, let
\[
dZ_{i,t}(u)
=
\mu_{i,t}(u)\,du
+
\sigma_{i,t}(u)\,dW_{i,t}(u),
\qquad u\in\mathcal U,
\]
where $Z_{i,t}(u)$ denotes the intraday log price, $\mu_{i,t}(u)$ is the local drift, $\sigma_{i,t}(u)$ is the spot volatility process, and $W_{i,t}(u)$ is a standard Brownian motion over the trading day. We define the curve component of the asset covariate as the latent spot volatility curve
\[
v_{i,t}(u)=\sigma_{i,t}(u),\qquad u\in\mathcal U.
\]
Thus, $v_{i,t}$ is not the raw high frequency return path, nor the raw price increment path. It is the latent intraday spot volatility curve that governs the local diffusion variation of the log price process.

At the population level, the intraday volatility state is therefore represented by the smooth curve $v_{i,t}=\sigma_{i,t}$. Since empirical intraday prices are observed only at discrete times, $v_{i,t}$ is not directly observed and must be replaced by a high frequency estimator. The theoretical object remains the latent curve $v_{i,t}$, while the empirical implementation uses an estimated grid representation $\widehat v_{i,t}$.

We impose regularity directly on the latent spot volatility curve. Let $\omega:[0,\infty)\to[0,\infty)$ be a common modulus of continuity satisfying $\omega(h)\to0$ as $h\downarrow0$. Define
\[
\mathcal V=
\{v\in C(\mathcal U): 0\le v\le \bar v,\ 
|v(u)-v(u')|\le \omega(|u-u'|)\}.
\]
Here the inequalities are understood pointwise for all $u,u'\in\mathcal U$, with $\bar v<\infty$. The nonnegativity is natural because $v$ is a volatility level. The boundedness and continuity conditions are imposed on the latent spot volatility process, not on the raw high frequency returns. In the empirical A share application, boundedness is also consistent with the finite variation induced by market trading rules such as price limits, although the exact limits may differ across stock types and market segments.

Let $\mathcal R=[-\bar R,\bar R]$ for some $\bar R<\infty$. The asset covariate is defined as the point curve object
\[
X_{i,t}=\bigl(R_{i,t},v_{i,t}\bigr)\in\mathcal X,
\qquad 
\mathcal X\equiv \mathcal R\times\mathcal V.
\]
We equip $\mathcal X$ with a product metric that separates the daily return coordinate and the intraday spot volatility coordinate. For $x=(r,v)$ and $x'=(r',v')$ in $\mathcal X$, define
\[
d_X(x,x')^2
:=
(r-r')^2+w\,\|v-v'\|_{L^2(\mathcal U)}^2,
\]
where $w>0$ controls the relative contribution of the curve component and
\[
\|v-v'\|_{L^2(\mathcal U)}^2
=
\int_{\mathcal U}
\{v(u)-v'(u)\}^2\,\mu_{\mathcal U}(du).
\]
This point curve construction enriches the information set used in the screening stage. Compared with a representation based only on returns, $X_{i,t}=(R_{i,t},v_{i,t})$ incorporates the full intraday spot volatility curve and therefore captures risk state information that is not visible from a single daily return.

In the data, intraday prices are recorded only at discrete sampling times. Hence the latent volatility curve $v_{i,t}$ is replaced by a finite grid estimator constructed from high frequency intraday returns. Since the MDS score only requires pairwise distances and empirical Fr\'echet means, we do not need to recover a continuous curve as an intermediate object. Instead, we use the finite grid estimator directly as a numerical representation of the latent intraday volatility state.

This simplification does not change the metric nature of the procedure. Under the $L^2$ metric used in this paper, the discretized curve distance is the usual Euclidean distance between grid vectors up to the grid weight, and the empirical Fr\'echet mean coincides with the componentwise average. This is a consequence of the particular metric adopted here, rather than a restriction of the MDS framework. With other object spaces or other metrics, the same construction can still be applied through distances and Fr\'echet means, even when these quantities no longer have Euclidean forms.

We now specify the finite grid estimator. Let $u_1<\cdots<u_J$ denote the five minute grid points on $\mathcal U$ at which the intraday volatility state is recorded. Let $\{s_\ell\}$ denote the one minute trading observation grid within day $t$, and define the log return between adjacent one minute trading observations by
\[
\tilde r_{i,t,\ell}
=
\log P_{i,t}(s_\ell)-\log P_{i,t}(s_{\ell-1}).
\]
The returns are computed over adjacent one minute trading observations within the same trading day, so 11:30 and 13:00 are adjacent in the trading observation sequence despite the midday calendar time break; overnight gaps are not used.

Let $\Delta$ denote the one minute sampling interval measured in the same intraday time scale as $\mathcal U$. For each five minute grid point $u_j$, let $\mathcal N_j$ be a local window of adjacent one minute return indices around $u_j$. In the empirical analysis, we use a symmetric window with two adjacent one minute returns on each side whenever available; at boundary points, the window is truncated to the available adjacent return indices. We estimate the spot volatility curve on the grid by
\[
\widehat v_{i,t,j}
=
\left(
\frac{1}{|\mathcal N_j|\Delta}
\sum_{\ell\in\mathcal N_j}
\tilde r_{i,t,\ell}^{\,2}
\right)^{1/2},
\qquad j=1,\ldots,J.
\]
In the empirical implementation, we use the one minute trading observation clock, so $\Delta=1$. 
This local realized estimator of spot volatility is motivated by the diffusion approximation
\[
\tilde r_{i,t,\ell}
\approx
\sigma_{i,t}(s_{\ell-1})\sqrt{\Delta}\,\varepsilon_{i,t,\ell},
\qquad 
\mathbb E(\varepsilon_{i,t,\ell}^2\mid\mathcal F_{s_{\ell-1}})=1.
\]
Thus, the local average of squared high frequency returns, after division by $\Delta$, estimates the local variance level $\sigma_{i,t}^2(u_j)$, and its square root estimates the spot volatility level $v_{i,t}(u_j)=\sigma_{i,t}(u_j)$.

To evaluate the $L^2(\mathcal U)$ term on the finite grid, we use an equally weighted discrete approximation,
\[
\int_{\mathcal U}
\{v_{i,t}(u)-v_{i,t'}(u)\}^2\,\mu_{\mathcal U}(du)
\approx
c_J\sum_{j=1}^{J}
\bigl(\widehat v_{i,t,j}-\widehat v_{i,t',j}\bigr)^2,
\]
where $c_J>0$ is a common grid weight. This gives the empirical distance
\[
d_X\!\left(\widehat X_{i,t},\widehat X_{i,t'}\right)^2
\approx
\left(R_{i,t}-R_{i,t'}\right)^2
+
w c_J\sum_{j=1}^{J}
\bigl(\widehat v_{i,t,j}-\widehat v_{i,t',j}\bigr)^2 .
\]
For implementation convenience, we use the normalized discrete distance
\[
d_X\!\left(\widehat X_{i,t},\widehat X_{i,t'}\right)^2
=
\left(R_{i,t}-R_{i,t'}\right)^2
+
\sum_{j=1}^{J}
\bigl(\widehat v_{i,t,j}-\widehat v_{i,t',j}\bigr)^2 .
\]
This normalization is kept fixed across all rolling windows and all methods in the empirical analysis. Other fixed positive weights for the curve component could also be used to adjust the relative scale of the intraday representation.

\subsubsection{High frequency Sharpe based value weighted target $Y$}\label{subsubsec:objects-Y}
To guide screening toward risk adjusted objectives, we construct a daily target return series by sorting stocks on a high frequency Sharpe ratio and then aggregating the selected stocks using market capitalization weights. This construction follows \citet{wang2022asset} and provides a one dimensional proxy for a risk adjusted investment opportunity. The resulting series serves as the target variable for ranking asset relevance in the screening step.

For each trading day $t$, let $P_{i,t}^{\mathrm{close}}$ denote the closing price of stock $i$ and define the daily return
\[
R_{i,t}\;=\;\frac{P_{i,t}^{\mathrm{close}}}{P_{i,t-1}^{\mathrm{close}}}-1.
\]
Let $R_{M,t}$ denote the daily return of the Shanghai Composite Index on day $t$. When intraday data are used, we divide day $t$ into $m$ equally spaced five minute intervals. Let $P_{i,t,j}$ denote the last observed price of stock $i$ in interval $j\in\{1,\ldots,J\}$. Define the intraday log return by
\[
r_{i,t,j}\;=\;\log P_{i,t,j}-\log P_{i,t,j-1},\qquad j=1,\ldots,J,
\]
where we set $P_{i,t,0}=P_{i,t-1}^{\mathrm{close}}$. Define the daily realized variance by $RV_{i,t}\;=\;\sum_{j=1}^J r_{i,t,j}^2$. We then define the individual high frequency Sharpe ratio by
\[
SR_{i,t}\;=\;\frac{R_{i,t}-R_{M,t}}{\sqrt{RV_{i,t}}}.
\]
If stock $i$ has no intraday transactions on day $t$ (so that $RV_{i,t}$ cannot be computed), we set $SR_{i,t}=-\infty$.

For each day $t$, rank $\{SR_{i,t}\}_{i=1}^p$ in descending order. Let $\mathcal{H}_{t}$ denote the set of stocks corresponding to the largest $q=\lfloor 0.1\,p\rfloor$ positive ratios. Let $MC_{i,t}$ denote the market capitalization of stock $i$ on day $t$ and define value weights
\[
w_{i,t}\;=\;\frac{MC_{i,t}}{\sum_{k\in\mathcal{H}_{t}} MC_{k,t}},\qquad i\in\mathcal{H}_{t}.
\]
The target return on day $t$ is the value weighted return of the selected set,
\[
Y_{t}\;=\;\sum_{i\in\mathcal{H}_{t}} w_{i,t}\,R_{i,t}.
\]

\begin{figure}[t]
	\centering
	\includegraphics[width=0.9\textwidth]{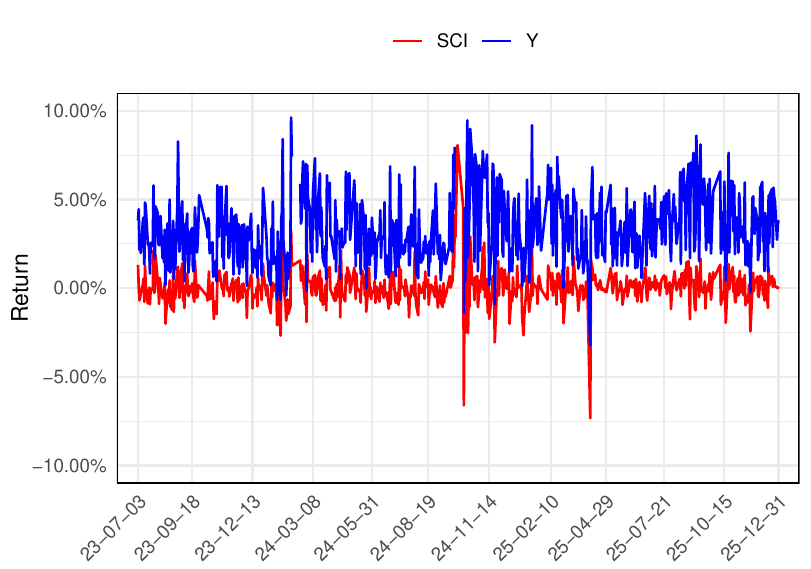}
	\caption{Daily returns of the target series $\{Y_t\}$ and the Shanghai Composite Index over the sample period.}
	\label{fig:target_vs_sci}
\end{figure}

Figure~\ref{fig:target_vs_sci} plots the daily target return series $\{Y_t\}$ together with the Shanghai Composite Index return series $\{R_{M,t}\}$ over the sample period. Two features are clear. First, the target series $\{Y_t\}$ is generally shifted upward relative to the index return series, which is consistent with the fact that it is formed from stocks with the highest daily high frequency Sharpe ratios. Second, the two series share pronounced market wide swings, which suggests that $\{Y_t\}$ captures common market shocks while retaining distinct variation induced by the Sharpe based sorting and aggregation. These features support the use of $\{Y_t\}$ as a risk adjusted target series for relevance ranking in the screening step.

The objects defined above form the basis of our screening method. In the next section, we introduce the dependence based relevance score for metric space valued objects and the resulting screening rule for ranking and selecting assets.

\subsection{Metric dependence screening}\label{subsec:mds}
We construct the first stage ranking score from a non-Euclidean correlation type coefficient defined on a metric space. 
Let $(X,Y)$ be a generic pair, where $X$ takes values in $(\mathcal X,d_X)$ and $Y$ is a real valued target taking values in $\mathcal Y\subset\mathbb R$, the support of $Y$.
The coefficient is based on an explained variation decomposition through Fr\'echet variance and is estimated through a partitioning scheme for the target. 
The population definition and the corresponding partition based estimator follow \citet{he2026frechetcorrelationcoefficientheterogeneous}. We summarize them below to fix notation and support the theoretical analysis in later sections.
	
\subsubsection{Population coefficient}\label{subsubsec:MDS-population}

Since $\mathcal R$ is bounded and $\mathcal V$ is uniformly bounded, $d_X$ is bounded on $\mathcal X$.
For $\mu\in\mathcal X$, define the Fr\'echet functional $F(\mu):=\mathbb E\,d_X^2(X,\mu)$.
Let $\mu_F:=\arg\min_{\mu\in\mathcal X}F(\mu)$ denote the global Fr\'echet mean, and define the corresponding global Fr\'echet variance by
\[
V_F:=\mathbb E\,d_X^2(X,\mu_F).
\]

To quantify the variation in $X$ explained by $Y$, define the conditional Fr\'echet functional $F(\mu\mid Y):=\mathbb E\!\left[d_X^2(X,\mu)\mid Y\right]$.
Let $\mu_F(Y)$ denote a measurable conditional Fr\'echet mean, with
$\mu_F(Y)\in\arg\min_{\mu\in\mathcal X}F(\mu\mid Y)$ almost surely.
The conditional residual variation is
\[
V(Y):=\inf_{\mu\in\mathcal X}\mathbb E\!\left[d_X^2(X,\mu)\mid Y\right]
=\mathbb E\!\left[d_X^2\bigl(X,\mu_F(Y)\bigr)\mid Y\right].
\]
The metric dependence coefficient is then defined as the explained variation ratio
\begin{equation}\label{eq:rho_population}
	\rho \;:=\; 1-\frac{\mathbb E\!\left[V(Y)\right]}{V_F}\ \in[0,1].
\end{equation}
By construction, $\rho$ is close to zero when conditioning on $Y$ does not reduce the metric dispersion of $X$. It becomes larger as $Y$ provides more information about the location of $X$ in $(\mathcal X,d_X)$.

\subsubsection{Sample estimation via partitioning the target}\label{subsubsec:MDS-estimation}
Let $\{(X_t,Y_t)\}_{t=1}^n$ be observations from a stationary time series.
Direct nonparametric estimation of $\mu_F(Y)$ is generally infeasible in a metric space. We therefore approximate conditioning on $Y$ by conditioning on a partition of its range, following \citet{he2026frechetcorrelationcoefficientheterogeneous}.
Let $\{\Omega_m\}_{m=1}^{M_n}$ be a measurable partition of the support of $Y$, and write $n_m:=\sum_{t=1}^n\mathbf 1\{Y_t\in\Omega_m\}$ and $\widehat p_m:=\frac{n_m}{n}$.

Define the empirical global Fr\'echet mean
\[
\widehat\mu_F\in\arg\min_{\mu\in\mathcal X}\frac1n\sum_{t=1}^n d_X^2(X_t,\mu),
\qquad
\widehat V_F:=\frac1n\sum_{t=1}^n d_X^2(X_t,\widehat\mu_F).
\]
For each cell $\Omega_m$ with $n_m>0$, define the empirical conditional Fr\'echet mean and variance by
\[
\widehat\mu_{F,m}\in\arg\min_{\mu\in\mathcal X}\frac1{n_m}\sum_{t:Y_t\in\Omega_m} d_{{X}}^2(X_t,\mu),
\quad
\widehat V_m:=\frac1{n_m}\sum_{t:Y_t\in\Omega_m} d_X^2(X_t,\widehat\mu_{F,m}).
\]
For cells with $n_m=0$, we set $\widehat p_m\widehat V_m=0$.
The plug-in estimator of \eqref{eq:rho_population} is
\begin{equation}\label{eq:rho_hat}
	\widehat\rho_n
	\;:=\;
	1-\frac{\sum_{m=1}^{M_n}\widehat p_m\,\widehat V_m}{\widehat V_F}.
\end{equation}
The numerator estimates the expected within cell residual variation, while the denominator estimates the unconditional variation. This preserves the explained variation interpretation.

\subsubsection{Asset level scores and screening rule}\label{subsubsec:MDS-screening}
For each asset $k\in\{1,\ldots,p\}$, apply \eqref{eq:rho_hat} to the sample $\{(X_{k,t},Y_t)\}_{t=1}^n$ under the metric $d_X$ defined in Section~\ref{subsubsec:objects-X}. This yields the asset level score
\[
\widehat\rho_n^{k}
\;:=\;
1-\frac{\sum_{m=1}^{M_n}\widehat p_m\,\widehat V_{m}^{k}}{\widehat V_F^{k}}.
\]
Here $\widehat V_F^{k}$ is the empirical Fr\'echet variance of $\{X_{k,t}\}_{t=1}^n$, and $\widehat V_m^{k}$ is the within cell empirical Fr\'echet variance computed from $\{X_{k,t}: Y_t\in\Omega_m\}$, as in Section~\ref{subsubsec:MDS-estimation}.
We rank $\{\widehat\rho_n^{k}\}_{k=1}^p$ in descending order and then screen assets by either a thresholding rule or a fixed size rule.

Let the active set be
\[
\mathcal A^\star:=\{1\le k\le p:\rho^{k}>0\},
\qquad s_n:=|\mathcal A^\star|,
\]
where $\rho^{k}$ is the population metric dependence coefficient in \eqref{eq:rho_population} for the pair $(X^{k},Y)$.
Given a threshold $\kappa n^{-\tau}$, define the screened set
\[
\widehat{\mathcal A}_n:=\Bigl\{1\le k\le p:\widehat\rho_n^{k}\ge \kappa\,n^{-\tau}\Bigr\}.
\]
Alternatively, given a target reduced dimension $d$, define the top $d$ screened set
\[
\widehat{\mathcal A}_n^\star
:=\Bigl\{k\in\{1,\ldots,p\}:\ \widehat\rho_n^{k}\ \text{is among the $d$ largest of }
\{\widehat\rho_n^{j}\}_{j=1}^p\Bigr\}.
\]
In our application, these screened sets provide a reduced asset universe for the second stage portfolio construction.

\section{Theoretical properties}\label{sec:MDS-theory}

This section presents theoretical properties of the proposed MDS procedure in dependent and high dimensional settings. We first state the assumptions and then present the main results on concentration, sure screening, and rank consistency.

We first state conditions on nondegeneracy, regularity of the point-curve space, and balancedness of the target partition.

\begin{assumption}
	\label{assumption:nonzero_variance}
	The global Fr\'echet variance $V_F=\mathbb E[d_X^2(X,\mu_F)]$ is strictly positive.
\end{assumption}

\begin{assumption}
	\label{assumption:holder-modulus}
	The common modulus of continuity in the definition of $\mathcal V$ satisfies
	$\omega(h)\le C_\omega h^\gamma$ for all $h\ge0$, for some constants
	$C_\omega<\infty$ and $\gamma\in(0,1]$.
\end{assumption}

\begin{assumption}
	\label{assumption:cell-prob}
	The number of cells satisfies $M_n\asymp n^a$ for some
	$0<a<\gamma/(1+3\gamma)$. The cell probabilities
	$p_m:=\Pr(Y\in\Omega_m)$ satisfy $p_m\asymp 1/M_n$ uniformly in $m$.
	Moreover, $\max_{1\le m\le M_n}\operatorname{diam}(\Omega_m)\lesssim M_n^{-1}$,
	where $\operatorname{diam}(\Omega_m):=\sup_{y,y'\in\Omega_m}|y-y'|$.
\end{assumption}

Assumptions~\ref{assumption:holder-modulus} and~\ref{assumption:cell-prob} play roles analogous to the smoothness and smoothing bandwidth conditions in kernel based screening methods such as \citet{wang2022asset}. Here the H\"older modulus controls the metric entropy of the point-curve space, while the cell probability and diameter conditions control the resolution of the target partition. Thus the partition size $M_n$ plays the role of a smoothing parameter: it must increase to reduce approximation bias, but not too quickly so that within cell estimation remains stable.

We next work under weak dependence and a smoothness condition for the conditional Fr\'echet functional. Assumption~\ref{assumption:holder-risk} controls the approximation error induced by partitioning the target. A smoothness condition of this type is also common in the Fr\'echet regression literature; see, for example, \citet{11475778}.

\begin{assumption}
	\label{assumption:mixing}
	The process $\{(X_i,Y_i)\}_{i\ge 1}$ is stationary and $\alpha$-mixing with coefficients $\{\alpha(k)\}_{k\ge1}$ satisfying $\alpha(k)\le e^{-c_0 k}$ for all $k\ge1$ and some constant $c_0>0$.
\end{assumption}

\begin{assumption}
	\label{assumption:holder-risk}
	There exist constants $\beta\in(0,1]$ and $L_R<\infty$ such that, for all $\mu\in\mathcal X$ and all $y,y'\in\mathcal Y$,
	$\big|\mathbb E[d_X^2(X,\mu)\mid Y=y]-\mathbb E[d_X^2(X,\mu)\mid Y=y']\big|\le L_R|y-y'|^\beta$.
\end{assumption}

Finally, we state the signal strength and separation conditions used to establish sure screening and rank consistency.

\begin{assumption}
	\label{assumption:signal}
	There exist constants $\kappa>0$ and
	$0\le \tau<\min\{a\beta,\frac{\gamma(1-3a)-a}{4\gamma+2}\}$
	such that, for all sufficiently large $n$,
	$\min_{k\in\mathcal A^\star}\rho^k \ge 2\kappa n^{-\tau}$.
\end{assumption}

\begin{assumption}
	\label{assumption:gap-and-growth}
	There exist constants $\kappa>0$ and
	$0\le \tau<\min\{a\beta,\frac{\gamma(1-3a)-a}{4\gamma+2}\}$
	such that, for all sufficiently large $n$,
	$\min_{k\in\mathcal A^\star}\rho^{k}
	-\max_{k\notin\mathcal A^\star}\rho^{k}
	\ge 2\kappa n^{-\tau}$,
	and the dimension $p=p_n$ satisfies $\log p_n=o(n^{1-3a-4\tau})$.
\end{assumption}

We now present the main theoretical results for the MDS coefficient and its empirical versions. The first result gives an exponential tail deviation bound for the estimator $\widehat\rho_n$ in \eqref{eq:rho_hat}. This bound provides nonasymptotic control and will be used to study the featurewise scores $\{\widehat\rho_n^k\}_{k=1}^p$ and their maxima in high dimensions.

\begin{theorem}\label{thm:exp-tail-rho}
	Under Assumptions~\ref{assumption:nonzero_variance}--\ref{assumption:holder-risk}, let $\rho$ and $\widehat\rho_n$ be defined in Sections~\ref{subsubsec:MDS-population} and~\ref{subsubsec:MDS-estimation}, respectively. Then for any $0\le \tau<\min\left\{a\beta,\frac{\gamma(1-3a)-a}{4\gamma+2}\right\}$, there exist constants $c,d>0$, independent of $n$, such that
	\[
	\Pr\bigl(|\widehat\rho_n-\rho|>dn^{-\tau}\bigr)
	=
	O\left(n^{a}\exp\left\{-cn^{1-3a-4\tau}\right\}\right) .
	\]
\end{theorem}

We next extend the concentration result to the featurewise dependence scores. In particular, we obtain an exponential tail control for the maximum deviation
$\max_{1\le k\le p}|\widehat\rho_n^k-\rho^k|$, which is the key step for establishing screening guarantees when $p=p_n$ can grow with $n$. Under the minimal signal condition in Assumption~\ref{assumption:signal}, we also obtain the sure screening property for the thresholded set $\widehat{\mathcal A}_n$.

\begin{theorem}\label{thm:max-p-rho}
	Under Assumptions~\ref{assumption:nonzero_variance}--\ref{assumption:holder-risk}, for any 
	$0\le \tau<\min\left\{a\beta,\frac{\gamma(1-3a)-a}{4\gamma+2}\right\}$, there exists a constant $c>0$  such that
	\[
	\Pr\Bigl(\max_{1\le k\le p}\bigl|\widehat\rho_n^k-\rho^k\bigr|>n^{-\tau}\Bigr)
	=
	O\left(
	pn^{a}\exp\bigl\{-cn^{1-3a-4\tau}\bigr\}
	\right).
	\]
	Moreover, if Assumption~\ref{assumption:signal} holds, then
	\[
	\Pr\bigl(\mathcal A^\star\subseteq \widehat{\mathcal A}_n\bigr)
	\ge1-O\left(
	s_nn^{a}\exp\bigl\{-cn^{1-3a-4\tau}\bigr\}
	\right),
	\]
	where $s_n:=|\mathcal A^\star|$.
\end{theorem}

Finally, we state a rank consistency result. Under an active-inactive separation condition and a mild growth restriction on $p_n$, the empirical scores asymptotically preserve the population ordering between active and inactive features. This result supports the use of the top $d$ rule $\widehat{\mathcal A}_n^\star$ when the threshold level $\kappa n^{-\tau}$ is not prespecified.

\begin{theorem}
	\label{thm:rank-consistency-rho}
	Under Assumptions~\ref{assumption:nonzero_variance}--\ref{assumption:holder-risk} and
	Assumption~\ref{assumption:gap-and-growth}, we have
	\[
	\liminf_{n\to\infty}
	\left(
	\min_{k\in\mathcal A^\star}\widehat\rho_n^{k}
	-\max_{k\notin\mathcal A^\star}\widehat\rho_n^{k}
	\right)\ge 0
	\quad\text{a.s.}
	\]
\end{theorem}

Theorem~\ref{thm:exp-tail-rho} establishes concentration for the explained variation coefficient under dependence and partition approximation. Theorem~\ref{thm:max-p-rho} extends this control to the high dimensional setting and yields sure screening for $\widehat{\mathcal A}_n$. Theorem~\ref{thm:rank-consistency-rho} shows that the induced ranking is asymptotically consistent and supports the practical use of the top $d$ screened set $\widehat{\mathcal A}_n^\star$.

\section{Simulation Studies}\label{sec:simulation}

In this section, we use Monte Carlo simulations to evaluate the finite sample performance of the proposed MDS procedure under serial dependence and ultrahigh dimensionality. We consider two settings. The first is a Euclidean setting, where both predictors and the target are Euclidean valued, so existing screening methods can be applied. The second is a non-Euclidean setting, where one variable is Euclidean valued and the other lies in a general metric space. This setting is used to examine whether MDS remains effective when the dependence involves object valued data.

The MDS framework applies to both settings. In the Euclidean setting, one dimensional Euclidean variables are special cases of metric valued objects. In the non-Euclidean setting, the object valued variable is treated as the metric valued random object, while the Euclidean variable provides the real valued conditioning variable used for partitioning. Hence the theoretical results in Section~\ref{sec:MDS-theory} are applicable to the simulation designs considered here. In the Euclidean setting, we compare MDS with SIS \citep{fan2008sure}, SIRS \citep{zhu2011model}, DC-SIS \citep{li2012feature}, and D-SEVIS \citep{wang2022asset}. In the non-Euclidean setting, these competitors are not applicable because they rely on Euclidean vector structure. We therefore focus on the finite sample performance of MDS itself.

To examine sensitivity to the screening size, we set $d_1=\lfloor n/\log(n)\rfloor$, $d_2=2\lfloor n/\log(n)\rfloor$, and $d_3=3\lfloor n/\log(n)\rfloor$, and report results for all three choices. For each configuration, we use 500 replications. We report two types of performance measures. First, $\mathcal P_s$ and $\mathcal P_a$ are reported, where $\mathcal P_s$ is the selection frequency of each active feature and $\mathcal P_a$ is the frequency with which the screened set contains all active features. Second, we report $\mathcal S$, the minimum screened set size required to include all active features, and summarize it by its 25\%, 50\%, and 75\% quantiles across replications. Since $\mathcal S$ is at least the number of active features, smaller values indicate better screening performance.

\subsection{Euclidean setting}\label{subsec:sim-euclid}
We first consider a Euclidean setting, where $X_i=(X_i^1,\ldots,X_i^p)^\top\in\mathbb R^p$ and the target $Y_i\in\mathbb R$ is scalar valued. We consider both linear and nonlinear signal structures. Specifically, we use the following three data generating models, which share the same active set $\mathcal A^\star=\{1,\,2,\,12,\,22\}$:
\begin{align*}
	\text{(1.a)}\quad
	Y &= 5X_{1} + 5X_{2} + 5X_{12} + 5X_{22} + \varepsilon,\\[2mm]
	\text{(1.b)}\quad
	Y &= 5X_{1} + 2X_{2} + 7\,\mathbf 1\{X_{12}<0\} + 5X_{22} + \varepsilon,\\[2mm]
	\text{(1.c)}\quad
	Y &= 5X_{1} + 2X_{2}^{2} + 5\sin(5X_{12}) + 5\max\{X_{22},0\} + \varepsilon,
\end{align*}
where $\{\varepsilon_i\}_{i=1}^n$ are i.i.d.\ $N(0,1)$ and independent of the predictor process. 

To reflect serial dependence and cross sectional correlation in high dimensions, we generate the predictor vector $X_i=(X_i^1,\ldots,X_i^p)^\top$ under two time series designs. In the first design, $\{X_i\}$ follows a vector AR(2) recursion
\[
X_i = 0.7\,X_{i-1}-0.4\,X_{i-2}+U_i,
\]
where $\{U_i\}$ are i.i.d.\ Gaussian innovations with mean zero and covariance matrix $\Sigma=(\Sigma_{jk})_{1\le j,k\le p}$ given by $\Sigma_{jk}=\sigma^{|j-k|}$. In the second design, $\{X_i\}$ follows a vector ARMA(1,1) recursion
\[
X_i = 0.5\,X_{i-1}+U_i-0.5\,U_{i-1},
\]
with the same innovation distribution and the same covariance matrix $\Sigma$. We set $n=200$, consider $p\in\{2000,5000\}$, and take $\sigma\in\{0.5,0.8\}$.

For each method, we compute the screened set for each $d\in\{d_1,d_2,d_3\}$ and summarize $\mathcal P_s$, $\mathcal P_a$, and the quantiles of $\mathcal S$ over 500 replications. The results are reported in Tables~\ref{tab:ar2-euclidean}--\ref{tab:example31-size}.

\begin{table}[H]
	\centering
	\caption{The proportions of $\mathcal{P}_s$ and $\mathcal{P}_a$ in Euclidean setting for AR(2).}
	\label{tab:ar2-euclidean}
	\resizebox{\textwidth}{!}{%
		\begin{tabular}{ll
				*{5}{c}
				*{5}{c}
				*{5}{c}
				*{5}{c}
				*{5}{c}}
			\toprule
			\multicolumn{2}{l}{}
			& \multicolumn{5}{c}{SIS}
			& \multicolumn{5}{c}{SIRS}
			& \multicolumn{5}{c}{DC-SIS}
			& \multicolumn{5}{c}{D-SEVIS}
			& \multicolumn{5}{c}{MDS} \\
			\cmidrule(lr){3-7}\cmidrule(lr){8-12}\cmidrule(lr){13-17}\cmidrule(lr){18-22}\cmidrule(lr){23-27}
			\multicolumn{2}{l}{}
			& \multicolumn{4}{c}{$\mathcal{P}_s$} & $\mathcal{P}_a$
			& \multicolumn{4}{c}{$\mathcal{P}_s$} & $\mathcal{P}_a$
			& \multicolumn{4}{c}{$\mathcal{P}_s$} & $\mathcal{P}_a$
			& \multicolumn{4}{c}{$\mathcal{P}_s$} & $\mathcal{P}_a$
			& \multicolumn{4}{c}{$\mathcal{P}_s$} & $\mathcal{P}_a$ \\
			\cmidrule(lr){3-6}\cmidrule(lr){8-11}\cmidrule(lr){13-16}\cmidrule(lr){18-21}\cmidrule(lr){23-26}
			Model & Size
			& $X_1$ & $X_2$ & $X_{12}$ & $X_{22}$ & ALL
			& $X_1$ & $X_2$ & $X_{12}$ & $X_{22}$ & ALL
			& $X_1$ & $X_2$ & $X_{12}$ & $X_{22}$ & ALL
			& $X_1$ & $X_2$ & $X_{12}$ & $X_{22}$ & ALL
			& $X_1$ & $X_2$ & $X_{12}$ & $X_{22}$ & ALL \\
			\midrule
			
			\multicolumn{27}{l}{Case 1: $p=2000\quad \sigma=0.5$} \\
			\midrule
			\multirow{3}{*}{(1.a)} & $d_1$ & 1.00 & 1.00 & 1.00 & 1.00 & \textbf{1.00} & 1.00 & 1.00 & 1.00 & 1.00 & \textbf{1.00} & 1.00 & 1.00 & 1.00 & 1.00 & \textbf{1.00} & 1.00 & 1.00 & 1.00 & 1.00 & \textbf{1.00} & 1.00 & 1.00 & 0.99 & 0.99 & \textbf{0.99} \\
			& $d_2$ & 1.00 & 1.00 & 1.00 & 1.00 & \textbf{1.00} & 1.00 & 1.00 & 1.00 & 1.00 & \textbf{1.00} & 1.00 & 1.00 & 1.00 & 1.00 & \textbf{1.00} & 1.00 & 1.00 & 1.00 & 1.00 & \textbf{1.00} & 1.00 & 1.00 & 1.00 & 1.00 & \textbf{0.99} \\
			& $d_3$ & 1.00 & 1.00 & 1.00 & 1.00 & \textbf{1.00} & 1.00 & 1.00 & 1.00 & 1.00 & \textbf{1.00} & 1.00 & 1.00 & 1.00 & 1.00 & \textbf{1.00} & 1.00 & 1.00 & 1.00 & 1.00 & \textbf{1.00} & 1.00 & 1.00 & 1.00 & 1.00 & \textbf{0.99} \\
			\multirow{3}{*}{(1.b)} & $d_1$ & 1.00 & 1.00 & 0.85 & 1.00 & \textbf{0.85} & 1.00 & 1.00 & 0.85 & 1.00 & \textbf{0.85} & 1.00 & 1.00 & 0.93 & 1.00 & \textbf{0.93} & 1.00 & 1.00 & 0.88 & 1.00 & \textbf{0.88} & 1.00 & 1.00 & 0.88 & 1.00 & \textbf{0.88}\\
			& $d_2$ & 1.00 & 1.00 & 0.92 & 1.00 & \textbf{0.92} & 1.00 & 1.00 & 0.89 & 1.00 & \textbf{0.89} & 1.00 & 1.00 & 0.96 & 1.00 & \textbf{0.96} & 1.00 & 1.00 & 0.94 & 1.00 & \textbf{0.94} & 1.00 & 1.00 & 0.94 & 1.00 & \textbf{0.94} \\
			& $d_3$ & 1.00 & 1.00 & 0.95 & 1.00 & \textbf{0.95} & 1.00 & 1.00 & 0.93 & 1.00 & \textbf{0.93} & 1.00 & 1.00 & 0.97 & 1.00 & \textbf{0.97} & 1.00 & 1.00 & 0.95 & 1.00 & \textbf{0.95} & 1.00 & 1.00 & 0.96 & 1.00 & \textbf{0.95} \\
			\multirow{3}{*}{(1.c)} & $d_1$ & 1.00 & 0.93 & 0.02 & 0.98 & \textbf{0.02} & 1.00 & 0.87 & 0.02 & 0.96 & \textbf{0.02} & 1.00 & 0.99 & 0.03 & 0.99 & \textbf{0.03} & 1.00 & 1.00 & 0.04 & 0.99 & \textbf{0.04} & 1.00 & 1.00 & 0.81 & 0.96 & \textbf{0.77} \\
			& $d_2$ & 1.00 & 0.96 & 0.05 & 0.98 & \textbf{0.04} & 1.00 & 0.92 & 0.04 & 0.99 & \textbf{0.04} & 1.00 & 1.00 & 0.07 & 1.00 & \textbf{0.06} & 1.00 & 1.00 & 0.07 & 0.99 & \textbf{0.07} & 1.00 & 1.00 & 0.87 & 0.98 & \textbf{0.86} \\
			& $d_3$ & 1.00 & 0.96 & 0.06 & 0.99 & \textbf{0.06} & 1.00 & 0.94 & 0.06 & 1.00 & \textbf{0.06} & 1.00 & 1.00 & 0.10 & 1.00 & \textbf{0.10} & 1.00 & 1.00 & 0.10 & 1.00 & \textbf{0.10} & 1.00 & 1.00 & 0.90 & 0.99 & \textbf{0.89} \\
			\midrule
			
			\multicolumn{27}{l}{Case 2: $p=2000\quad \sigma=0.8$} \\
			\midrule
			\multirow{3}{*}{(1.a)} & $d_1$ & 1.00 & 1.00 & 1.00 & 1.00 & \textbf{1.00} & 1.00 & 1.00 & 1.00 & 1.00 & \textbf{1.00} & 1.00 & 1.00 & 1.00 & 1.00 & \textbf{1.00} & 1.00 & 1.00 & 1.00 & 1.00 & \textbf{1.00} & 1.00 & 1.00 & 1.00 & 0.99 & \textbf{0.99} \\
			& $d_2$ & 1.00 & 1.00 & 1.00 & 1.00 & \textbf{1.00} & 1.00 & 1.00 & 1.00 & 1.00 & \textbf{1.00} & 1.00 & 1.00 & 1.00 & 1.00 & \textbf{1.00} & 1.00 & 1.00 & 1.00 & 1.00 & \textbf{1.00} & 1.00 & 1.00 & 1.00 & 1.00 & \textbf{1.00} \\
			& $d_3$ & 1.00 & 1.00 & 1.00 & 1.00 & \textbf{1.00} & 1.00 & 1.00 & 1.00 & 1.00 & \textbf{1.00} & 1.00 & 1.00 & 1.00 & 1.00 & \textbf{1.00} & 1.00 & 1.00 & 1.00 & 1.00 & \textbf{1.00} & 1.00 & 1.00 & 1.00 & 1.00 & \textbf{1.00} \\
			\multirow{3}{*}{(1.b)} & $d_1$ & 1.00 & 1.00 & 0.22 & 1.00 & \textbf{0.22} & 1.00 & 1.00 & 0.22 & 1.00 & \textbf{0.22} & 1.00 & 1.00 & 0.38 & 1.00 & \textbf{0.38} & 1.00 & 1.00 & 0.36 & 1.00 & \textbf{0.36} & 1.00 & 1.00 & 0.43 & 1.00 & \textbf{0.43} \\
			& $d_2$ & 1.00 & 1.00 & 0.35 & 1.00 & \textbf{0.35} & 1.00 & 1.00 & 0.34 & 1.00 & \textbf{0.34} & 1.00 & 1.00 & 0.55 & 1.00 & \textbf{0.55} & 1.00 & 1.00 & 0.48 & 1.00 & \textbf{0.48} & 1.00 & 1.00 & 0.51 & 1.00 & \textbf{0.51} \\
			& $d_3$ & 1.00 & 1.00 & 0.43 & 1.00 & \textbf{0.43} & 1.00 & 1.00 & 0.40 & 1.00 & \textbf{0.40} & 1.00 & 1.00 & 0.61 & 1.00 & \textbf{0.61} & 1.00 & 1.00 & 0.56 & 1.00 & \textbf{0.56} & 1.00 & 1.00 & 0.60 & 1.00 & \textbf{0.60} \\
			\multirow{3}{*}{(1.c)} & $d_1$ & 1.00 & 1.00 & 0.09 & 0.98 & \textbf{0.09} & 1.00 & 1.00 & 0.07 & 0.99 & \textbf{0.07} & 1.00 & 1.00 & 0.14 & 1.00 & \textbf{0.14} & 1.00 & 1.00 & 0.09 & 0.99 & \textbf{0.09} & 1.00 & 1.00 & 0.87 & 0.97 & \textbf{0.84} \\
			& $d_2$ & 1.00 & 1.00 & 0.19 & 1.00 & \textbf{0.19} & 1.00 & 1.00 & 0.15 & 1.00 & \textbf{0.15} & 1.00 & 1.00 & 0.24 & 1.00 & \textbf{0.24} & 1.00 & 1.00 & 0.15 & 0.99 & \textbf{0.15} & 1.00 & 1.00 & 0.92 & 0.98 & \textbf{0.90} \\
			& $d_3$ & 1.00 & 1.00 & 0.23 & 1.00 & \textbf{0.23} & 1.00 & 1.00 & 0.21 & 1.00 & \textbf{0.21} & 1.00 & 1.00 & 0.29 & 1.00 & \textbf{0.29} & 1.00 & 1.00 & 0.21 & 1.00 & \textbf{0.21} & 1.00 & 1.00 & 0.95 & 0.99 & \textbf{0.94} \\
			\midrule
			
			\multicolumn{27}{l}{Case 3: $p=5000\quad \sigma=0.5$} \\
			\midrule
			\multirow{3}{*}{(1.a)} & $d_1$ & 1.00 & 1.00 & 1.00 & 1.00 & \textbf{1.00} & 1.00 & 1.00 & 1.00 & 1.00 & \textbf{1.00} & 1.00 & 1.00 & 1.00 & 1.00 & \textbf{1.00} & 1.00 & 1.00 & 1.00 & 1.00 & \textbf{1.00} & 1.00 & 1.00 & 0.99 & 0.99 & \textbf{0.98} \\
			& $d_2$ & 1.00 & 1.00 & 1.00 & 1.00 & \textbf{1.00} & 1.00 & 1.00 & 1.00 & 1.00 & \textbf{1.00} & 1.00 & 1.00 & 1.00 & 1.00 & \textbf{1.00} & 1.00 & 1.00 & 1.00 & 1.00 & \textbf{1.00} & 1.00 & 1.00 & 0.99 & 0.99 & \textbf{0.99} \\
			& $d_3$ & 1.00 & 1.00 & 1.00 & 1.00 & \textbf{1.00} & 1.00 & 1.00 & 1.00 & 1.00 & \textbf{1.00} & 1.00 & 1.00 & 1.00 & 1.00 & \textbf{1.00} & 1.00 & 1.00 & 1.00 & 1.00 & \textbf{1.00} & 1.00 & 1.00 & 1.00 & 1.00 & \textbf{0.99} \\
			\multirow{3}{*}{(1.b)} & $d_1$ & 1.00 & 1.00 & 0.78 & 1.00 & \textbf{0.78} & 1.00 & 1.00 & 0.76 & 1.00 & \textbf{0.76} & 1.00 & 1.00 & 0.87 & 1.00 & \textbf{0.87} & 1.00 & 1.00 & 0.81 & 1.00 & \textbf{0.81} & 1.00 & 1.00 & 0.81 & 1.00 & \textbf{0.81} \\
			& $d_2$ & 1.00 & 1.00 & 0.86 & 1.00 & \textbf{0.86} & 1.00 & 1.00 & 0.84 & 1.00 & \textbf{0.84} & 1.00 & 1.00 & 0.91 & 1.00 & \textbf{0.91} & 1.00 & 1.00 & 0.86 & 1.00 & \textbf{0.86} & 1.00 & 1.00 & 0.87 & 1.00 & \textbf{0.87} \\
			& $d_3$ & 1.00 & 1.00 & 0.89 & 1.00 & \textbf{0.89} & 1.00 & 1.00 & 0.87 & 1.00 & \textbf{0.87} & 1.00 & 1.00 & 0.94 & 1.00 & \textbf{0.94} & 1.00 & 1.00 & 0.89 & 1.00 & \textbf{0.89} & 1.00 & 1.00 & 0.88 & 1.00 & \textbf{0.88} \\
			\multirow{3}{*}{(1.c)} & $d_1$ & 1.00 & 0.92 & 0.02 & 0.97 & \textbf{0.01} & 1.00 & 0.85 & 0.01 & 0.96 & \textbf{0.00} & 1.00 & 0.99 & 0.01 & 0.98 & \textbf{0.01} & 1.00 & 1.00 & 0.01 & 0.98 & \textbf{0.01} & 1.00 & 0.99 & 0.71 & 0.95 & \textbf{0.67} \\
			& $d_2$ & 1.00 & 0.94 & 0.02 & 0.99 & \textbf{0.02} & 1.00 & 0.90 & 0.02 & 0.98 & \textbf{0.01} & 1.00 & 0.99 & 0.04 & 0.99 & \textbf{0.04} & 1.00 & 1.00 & 0.04 & 0.99 & \textbf{0.04} & 1.00 & 1.00 & 0.80 & 0.98 & \textbf{0.78} \\
			& $d_3$ & 1.00 & 0.94 & 0.03 & 0.99 & \textbf{0.02} & 1.00 & 0.92 & 0.03 & 0.99 & \textbf{0.02} & 1.00 & 0.99 & 0.05 & 1.00 & \textbf{0.05} & 1.00 & 1.00 & 0.05 & 0.99 & \textbf{0.05} & 1.00 & 1.00 & 0.82 & 0.98 & \textbf{0.80} \\
			\midrule
			
			\multicolumn{27}{l}{Case 4: $p=5000\quad \sigma=0.8$} \\
			\midrule
			\multirow{3}{*}{(1.a)} & $d_1$ & 1.00 & 1.00 & 1.00 & 1.00 & \textbf{1.00} & 1.00 & 1.00 & 1.00 & 1.00 & \textbf{1.00} & 1.00 & 1.00 & 1.00 & 0.99 & \textbf{0.99} & 1.00 & 1.00 & 1.00 & 0.99 & \textbf{0.99} & 1.00 & 1.00 & 1.00 & 0.97 & \textbf{0.97} \\
			& $d_2$ & 1.00 & 1.00 & 1.00 & 1.00 & \textbf{1.00} & 1.00 & 1.00 & 1.00 & 1.00 & \textbf{1.00} & 1.00 & 1.00 & 1.00 & 1.00 & \textbf{1.00} & 1.00 & 1.00 & 1.00 & 1.00 & \textbf{1.00} & 1.00 & 1.00 & 1.00 & 0.98 & \textbf{0.98} \\
			& $d_3$ & 1.00 & 1.00 & 1.00 & 1.00 & \textbf{1.00} & 1.00 & 1.00 & 1.00 & 1.00 & \textbf{1.00} & 1.00 & 1.00 & 1.00 & 1.00 & \textbf{1.00} & 1.00 & 1.00 & 1.00 & 1.00 & \textbf{1.00} & 1.00 & 1.00 & 1.00 & 0.99 & \textbf{0.99} \\
			\multirow{3}{*}{(1.b)} & $d_1$ & 1.00 & 1.00 & 0.17 & 1.00 & \textbf{0.17} & 1.00 & 1.00 & 0.16 & 1.00 & \textbf{0.16} & 1.00 & 1.00 & 0.33 & 1.00 & \textbf{0.33} & 1.00 & 1.00 & 0.28 & 1.00 & \textbf{0.28} & 1.00 & 1.00 & 0.34 & 1.00 & \textbf{0.34} \\
			& $d_2$ & 1.00 & 1.00 & 0.28 & 1.00 & \textbf{0.28} & 1.00 & 1.00 & 0.27 & 1.00 & \textbf{0.27} & 1.00 & 1.00 & 0.42 & 1.00 & \textbf{0.42} & 1.00 & 1.00 & 0.38 & 1.00 & \textbf{0.38} & 1.00 & 1.00 & 0.46 & 1.00 & \textbf{0.46} \\
			& $d_3$ & 1.00 & 1.00 & 0.32 & 1.00 & \textbf{0.32} & 1.00 & 1.00 & 0.33 & 1.00 & \textbf{0.33} & 1.00 & 1.00 & 0.49 & 1.00 & \textbf{0.49} & 1.00 & 1.00 & 0.44 & 1.00 & \textbf{0.44} & 1.00 & 1.00 & 0.54 & 1.00 & \textbf{0.54} \\
			\multirow{3}{*}{(1.c)} & $d_1$ & 1.00 & 1.00 & 0.03 & 0.95 & \textbf{0.03} & 1.00 & 1.00 & 0.02 & 0.96 & \textbf{0.02} & 1.00 & 1.00 & 0.05 & 0.99 & \textbf{0.05} & 1.00 & 1.00 & 0.03 & 0.97 & \textbf{0.03} & 1.00 & 1.00 & 0.77 & 0.93 & \textbf{0.72} \\
			& $d_2$ & 1.00 & 1.00 & 0.08 & 0.98 & \textbf{0.08} & 1.00 & 1.00 & 0.06 & 0.99 & \textbf{0.06} & 1.00 & 1.00 & 0.09 & 1.00 & \textbf{0.09} & 1.00 & 1.00 & 0.06 & 0.99 & \textbf{0.06} & 1.00 & 1.00 & 0.85 & 0.96 & \textbf{0.82} \\
			& $d_3$ & 1.00 & 1.00 & 0.11 & 0.99 & \textbf{0.11} & 1.00 & 1.00 & 0.08 & 0.99 & \textbf{0.08} & 1.00 & 1.00 & 0.14 & 1.00 & \textbf{0.14} & 1.00 & 1.00 & 0.09 & 0.99 & \textbf{0.09} & 1.00 & 1.00 & 0.89 & 0.97 & \textbf{0.87} \\
			\bottomrule
		\end{tabular}%
	}
	\vspace{2mm}
	
	\begin{minipage}{0.98\textwidth}
		\footnotesize
		\textbf{Notes:} Numbers stand for proportions over 500 replications. $\mathcal{P}_s$ is the proportion
		that each true active feature is selected. $\mathcal{P}_a$ is the proportion that all true predictors are selected.
	\end{minipage}
\end{table}

\begin{table}[H]
	\centering
	\caption{The proportions of $\mathcal{P}_s$ and $\mathcal{P}_a$ in Euclidean setting for ARMA(1,1).}
	\label{tab:arma-euclidean}
	\resizebox{\textwidth}{!}{%
		\begin{tabular}{ll
				*{5}{c}
				*{5}{c}
				*{5}{c}
				*{5}{c}
				*{5}{c}}
			\toprule
			\multicolumn{2}{l}{}
			& \multicolumn{5}{c}{SIS}
			& \multicolumn{5}{c}{SIRS}
			& \multicolumn{5}{c}{DC-SIS}
			& \multicolumn{5}{c}{D-SEVIS}
			& \multicolumn{5}{c}{MDS} \\
			\cmidrule(lr){3-7}\cmidrule(lr){8-12}\cmidrule(lr){13-17}\cmidrule(lr){18-22}\cmidrule(lr){23-27}
			\multicolumn{2}{l}{}
			& \multicolumn{4}{c}{$\mathcal{P}_s$} & $\mathcal{P}_a$
			& \multicolumn{4}{c}{$\mathcal{P}_s$} & $\mathcal{P}_a$
			& \multicolumn{4}{c}{$\mathcal{P}_s$} & $\mathcal{P}_a$
			& \multicolumn{4}{c}{$\mathcal{P}_s$} & $\mathcal{P}_a$
			& \multicolumn{4}{c}{$\mathcal{P}_s$} & $\mathcal{P}_a$ \\
			\cmidrule(lr){3-6}\cmidrule(lr){8-11}\cmidrule(lr){13-16}\cmidrule(lr){18-21}\cmidrule(lr){23-26}
			Model & Size
			& $X_1$ & $X_2$ & $X_{12}$ & $X_{22}$ & ALL
			& $X_1$ & $X_2$ & $X_{12}$ & $X_{22}$ & ALL
			& $X_1$ & $X_2$ & $X_{12}$ & $X_{22}$ & ALL
			& $X_1$ & $X_2$ & $X_{12}$ & $X_{22}$ & ALL
			& $X_1$ & $X_2$ & $X_{12}$ & $X_{22}$ & ALL \\
			\midrule
			
			\multicolumn{27}{l}{Case 1: $p=2000\quad \sigma=0.5$} \\
			\midrule
			\multirow{3}{*}{(1.a)} & $d_1$ & 1.00 & 1.00 & 1.00 & 1.00 & \textbf{1.00} & 1.00 & 1.00 & 1.00 & 1.00 & \textbf{1.00} & 1.00 & 1.00 & 1.00 & 1.00 & \textbf{1.00} & 1.00 & 1.00 & 1.00 & 1.00 & \textbf{1.00} & 1.00 & 1.00 & 1.00 & 0.99 & \textbf{0.99} \\
			& $d_2$ & 1.00 & 1.00 & 1.00 & 1.00 & \textbf{1.00} & 1.00 & 1.00 & 1.00 & 1.00 & \textbf{1.00} & 1.00 & 1.00 & 1.00 & 1.00 & \textbf{1.00} & 1.00 & 1.00 & 1.00 & 1.00 & \textbf{1.00} & 1.00 & 1.00 & 1.00 & 1.00 & \textbf{1.00} \\
			& $d_3$ & 1.00 & 1.00 & 1.00 & 1.00 & \textbf{1.00} & 1.00 & 1.00 & 1.00 & 1.00 & \textbf{1.00} & 1.00 & 1.00 & 1.00 & 1.00 & \textbf{1.00} & 1.00 & 1.00 & 1.00 & 1.00 & \textbf{1.00} & 1.00 & 1.00 & 1.00 & 1.00 & \textbf{1.00} \\
			\multirow{3}{*}{(1.b)} & $d_1$ & 1.00 & 1.00 & 0.99 & 1.00 & \textbf{0.99} & 1.00 & 1.00 & 0.98 & 1.00 & \textbf{0.98} & 1.00 & 1.00 & 1.00 & 1.00 & \textbf{1.00} & 1.00 & 1.00 & 0.99 & 1.00 & \textbf{0.99} & 1.00 & 1.00 & 0.98 & 1.00 & \textbf{0.98} \\
			& $d_2$ & 1.00 & 1.00 & 1.00 & 1.00 & \textbf{1.00} & 1.00 & 1.00 & 1.00 & 1.00 & \textbf{1.00} & 1.00 & 1.00 & 1.00 & 1.00 & \textbf{1.00} & 1.00 & 1.00 & 0.99 & 1.00 & \textbf{0.99} & 1.00 & 1.00 & 0.99 & 1.00 & \textbf{0.99} \\
			& $d_3$ & 1.00 & 1.00 & 1.00 & 1.00 & \textbf{1.00} & 1.00 & 1.00 & 1.00 & 1.00 & \textbf{1.00} & 1.00 & 1.00 & 1.00 & 1.00 & \textbf{1.00} & 1.00 & 1.00 & 1.00 & 1.00 & \textbf{1.00} & 1.00 & 1.00 & 0.99 & 1.00 & \textbf{0.99} \\
			\multirow{3}{*}{(1.c)} & $d_1$ & 1.00 & 0.98 & 0.01 & 1.00 & \textbf{0.01} & 1.00 & 0.95 & 0.01 & 1.00 & \textbf{0.00} & 1.00 & 1.00 & 0.05 & 1.00 & \textbf{0.05} & 1.00 & 1.00 & 0.10 & 1.00 & \textbf{0.10} & 1.00 & 1.00 & 0.98 & 0.98 & \textbf{0.96} \\
			& $d_2$ & 1.00 & 0.98 & 0.03 & 1.00 & \textbf{0.03} & 1.00 & 0.97 & 0.03 & 1.00 & \textbf{0.02} & 1.00 & 1.00 & 0.12 & 1.00 & \textbf{0.12} & 1.00 & 1.00 & 0.16 & 1.00 & \textbf{0.16} & 1.00 & 1.00 & 0.99 & 1.00 & \textbf{0.98} \\
			& $d_3$ & 1.00 & 0.99 & 0.05 & 1.00 & \textbf{0.05} & 1.00 & 0.98 & 0.04 & 1.00 & \textbf{0.04} & 1.00 & 1.00 & 0.22 & 1.00 & \textbf{0.22} & 1.00 & 1.00 & 0.23 & 1.00 & \textbf{0.23} & 1.00 & 1.00 & 0.99 & 1.00 & \textbf{0.99} \\
			\midrule
			
			\multicolumn{27}{l}{Case 2: $p=2000\quad \sigma=0.8$} \\
			\midrule
			\multirow{3}{*}{(1.a)} & $d_1$ & 1.00 & 1.00 & 1.00 & 1.00 & \textbf{1.00} & 1.00 & 1.00 & 1.00 & 1.00 & \textbf{1.00} & 1.00 & 1.00 & 1.00 & 1.00 & \textbf{1.00} & 1.00 & 1.00 & 1.00 & 1.00 & \textbf{1.00} & 1.00 & 1.00 & 1.00 & 1.00 & \textbf{1.00} \\
			& $d_2$ & 1.00 & 1.00 & 1.00 & 1.00 & \textbf{1.00} & 1.00 & 1.00 & 1.00 & 1.00 & \textbf{1.00} & 1.00 & 1.00 & 1.00 & 1.00 & \textbf{1.00} & 1.00 & 1.00 & 1.00 & 1.00 & \textbf{1.00} & 1.00 & 1.00 & 1.00 & 1.00 & \textbf{1.00} \\
			& $d_3$ & 1.00 & 1.00 & 1.00 & 1.00 & \textbf{1.00} & 1.00 & 1.00 & 1.00 & 1.00 & \textbf{1.00} & 1.00 & 1.00 & 1.00 & 1.00 & \textbf{1.00} & 1.00 & 1.00 & 1.00 & 1.00 & \textbf{1.00} & 1.00 & 1.00 & 1.00 & 1.00 & \textbf{1.00} \\
			\multirow{3}{*}{(1.b)} & $d_1$ & 1.00 & 1.00 & 0.53 & 1.00 & \textbf{0.53} & 1.00 & 1.00 & 0.50 & 1.00 & \textbf{0.50} & 1.00 & 1.00 & 0.74 & 1.00 & \textbf{0.74} & 1.00 & 1.00 & 0.64 & 1.00 & \textbf{0.64} & 1.00 & 1.00 & 0.71 & 1.00 & \textbf{0.71} \\
			& $d_2$ & 1.00 & 1.00 & 0.67 & 1.00 & \textbf{0.67} & 1.00 & 1.00 & 0.62 & 1.00 & \textbf{0.62} & 1.00 & 1.00 & 0.84 & 1.00 & \textbf{0.84} & 1.00 & 1.00 & 0.77 & 1.00 & \textbf{0.77} & 1.00 & 1.00 & 0.83 & 1.00 & \textbf{0.83} \\
			& $d_3$ & 1.00 & 1.00 & 0.73 & 1.00 & \textbf{0.73} & 1.00 & 1.00 & 0.71 & 1.00 & \textbf{0.71} & 1.00 & 1.00 & 0.88 & 1.00 & \textbf{0.88} & 1.00 & 1.00 & 0.83 & 1.00 & \textbf{0.83} & 1.00 & 1.00 & 0.87 & 1.00 & \textbf{0.87} \\
			\multirow{3}{*}{(1.c)} & $d_1$ & 1.00 & 1.00 & 0.10 & 0.99 & \textbf{0.10} & 1.00 & 1.00 & 0.09 & 0.99 & \textbf{0.08} & 1.00 & 1.00 & 0.20 & 1.00 & \textbf{0.20} & 1.00 & 1.00 & 0.17 & 0.99 & \textbf{0.17} & 1.00 & 1.00 & 0.98 & 0.98 & \textbf{0.96} \\
			& $d_2$ & 1.00 & 1.00 & 0.19 & 1.00 & \textbf{0.19} & 1.00 & 1.00 & 0.18 & 1.00 & \textbf{0.18} & 1.00 & 1.00 & 0.35 & 1.00 & \textbf{0.35} & 1.00 & 1.00 & 0.31 & 1.00 & \textbf{0.31} & 1.00 & 1.00 & 1.00 & 0.98 & \textbf{0.98} \\
			& $d_3$ & 1.00 & 1.00 & 0.24 & 1.00 & \textbf{0.24} & 1.00 & 1.00 & 0.24 & 1.00 & \textbf{0.24} & 1.00 & 1.00 & 0.46 & 1.00 & \textbf{0.46} & 1.00 & 1.00 & 0.39 & 1.00 & \textbf{0.39} & 1.00 & 1.00 & 1.00 & 0.99 & \textbf{0.99} \\
			\midrule
			
			\multicolumn{27}{l}{Case 3: $p=5000\quad \sigma=0.5$} \\
			\midrule
			\multirow{3}{*}{(1.a)} & $d_1$ & 1.00 & 1.00 & 1.00 & 1.00 & \textbf{1.00} & 1.00 & 1.00 & 1.00 & 1.00 & \textbf{1.00} & 1.00 & 1.00 & 1.00 & 1.00 & \textbf{1.00} & 1.00 & 1.00 & 1.00 & 1.00 & \textbf{1.00} & 1.00 & 1.00 & 1.00 & 1.00 & \textbf{1.00} \\
			& $d_2$ & 1.00 & 1.00 & 1.00 & 1.00 & \textbf{1.00} & 1.00 & 1.00 & 1.00 & 1.00 & \textbf{1.00} & 1.00 & 1.00 & 1.00 & 1.00 & \textbf{1.00} & 1.00 & 1.00 & 1.00 & 1.00 & \textbf{1.00} & 1.00 & 1.00 & 1.00 & 1.00 & \textbf{1.00} \\
			& $d_3$ & 1.00 & 1.00 & 1.00 & 1.00 & \textbf{1.00} & 1.00 & 1.00 & 1.00 & 1.00 & \textbf{1.00} & 1.00 & 1.00 & 1.00 & 1.00 & \textbf{1.00} & 1.00 & 1.00 & 1.00 & 1.00 & \textbf{1.00} & 1.00 & 1.00 & 1.00 & 1.00 & \textbf{1.00} \\
			\multirow{3}{*}{(1.b)} & $d_1$ & 1.00 & 1.00 & 0.97 & 1.00 & \textbf{0.97} & 1.00 & 1.00 & 0.96 & 1.00 & \textbf{0.96} & 1.00 & 1.00 & 0.99 & 1.00 & \textbf{0.99} & 1.00 & 1.00 & 0.97 & 1.00 & \textbf{0.97} & 1.00 & 1.00 & 0.95 & 1.00 & \textbf{0.95} \\
			& $d_2$ & 1.00 & 1.00 & 0.99 & 1.00 & \textbf{0.99} & 1.00 & 1.00 & 0.98 & 1.00 & \textbf{0.98} & 1.00 & 1.00 & 1.00 & 1.00 & \textbf{1.00} & 1.00 & 1.00 & 0.98 & 1.00 & \textbf{0.98} & 1.00 & 1.00 & 0.98 & 1.00 & \textbf{0.98} \\
			& $d_3$ & 1.00 & 1.00 & 1.00 & 1.00 & \textbf{1.00} & 1.00 & 1.00 & 0.99 & 1.00 & \textbf{0.99} & 1.00 & 1.00 & 1.00 & 1.00 & \textbf{1.00} & 1.00 & 1.00 & 0.98 & 1.00 & \textbf{0.98} & 1.00 & 1.00 & 0.99 & 1.00 & \textbf{0.99} \\
			\multirow{3}{*}{(1.c)} & $d_1$ & 1.00 & 0.96 & 0.00 & 1.00 & \textbf{0.00} & 1.00 & 0.93 & 0.01 & 0.99 & \textbf{0.01} & 1.00 & 1.00 & 0.03 & 1.00 & \textbf{0.03} & 1.00 & 1.00 & 0.06 & 0.99 & \textbf{0.06} & 1.00 & 1.00 & 0.96 & 0.97 & \textbf{0.93} \\
			& $d_2$ & 1.00 & 0.98 & 0.02 & 1.00 & \textbf{0.02} & 1.00 & 0.95 & 0.01 & 0.99 & \textbf{0.01} & 1.00 & 1.00 & 0.06 & 1.00 & \textbf{0.06} & 1.00 & 1.00 & 0.09 & 1.00 & \textbf{0.09} & 1.00 & 1.00 & 0.98 & 0.98 & \textbf{0.97} \\
			& $d_3$ & 1.00 & 0.99 & 0.03 & 1.00 & \textbf{0.03} & 1.00 & 0.96 & 0.03 & 1.00 & \textbf{0.03} & 1.00 & 1.00 & 0.10 & 1.00 & \textbf{0.10} & 1.00 & 1.00 & 0.12 & 1.00 & \textbf{0.12} & 1.00 & 1.00 & 0.99 & 0.99 & \textbf{0.98} \\
			\midrule
			
			\multicolumn{27}{l}{Case 4: $p=5000\quad \sigma=0.8$} \\
			\midrule
			\multirow{3}{*}{(1.a)} & $d_1$ & 1.00 & 1.00 & 1.00 & 1.00 & \textbf{1.00} & 1.00 & 1.00 & 1.00 & 1.00 & \textbf{1.00} & 1.00 & 1.00 & 1.00 & 1.00 & \textbf{1.00} & 1.00 & 1.00 & 1.00 & 1.00 & \textbf{1.00} & 1.00 & 1.00 & 1.00 & 1.00 & \textbf{1.00} \\
			& $d_2$ & 1.00 & 1.00 & 1.00 & 1.00 & \textbf{1.00} & 1.00 & 1.00 & 1.00 & 1.00 & \textbf{1.00} & 1.00 & 1.00 & 1.00 & 1.00 & \textbf{1.00} & 1.00 & 1.00 & 1.00 & 1.00 & \textbf{1.00} & 1.00 & 1.00 & 1.00 & 1.00 & \textbf{1.00} \\
			& $d_3$ & 1.00 & 1.00 & 1.00 & 1.00 & \textbf{1.00} & 1.00 & 1.00 & 1.00 & 1.00 & \textbf{1.00} & 1.00 & 1.00 & 1.00 & 1.00 & \textbf{1.00} & 1.00 & 1.00 & 1.00 & 1.00 & \textbf{1.00} & 1.00 & 1.00 & 1.00 & 1.00 & \textbf{1.00} \\
			\multirow{3}{*}{(1.b)} & $d_1$ & 1.00 & 1.00 & 0.42 & 1.00 & \textbf{0.42} & 1.00 & 1.00 & 0.40 & 1.00 & \textbf{0.40} & 1.00 & 1.00 & 0.67 & 1.00 & \textbf{0.67} & 1.00 & 1.00 & 0.58 & 1.00 & \textbf{0.58} & 1.00 & 1.00 & 0.62 & 1.00 & \textbf{0.62} \\
			& $d_2$ & 1.00 & 1.00 & 0.56 & 1.00 & \textbf{0.56} & 1.00 & 1.00 & 0.52 & 1.00 & \textbf{0.52} & 1.00 & 1.00 & 0.78 & 1.00 & \textbf{0.78} & 1.00 & 1.00 & 0.67 & 1.00 & \textbf{0.67} & 1.00 & 1.00 & 0.72 & 1.00 & \textbf{0.72} \\
			& $d_3$ & 1.00 & 1.00 & 0.63 & 1.00 & \textbf{0.63} & 1.00 & 1.00 & 0.59 & 1.00 & \textbf{0.59} & 1.00 & 1.00 & 0.83 & 1.00 & \textbf{0.83} & 1.00 & 1.00 & 0.73 & 1.00 & \textbf{0.73} & 1.00 & 1.00 & 0.78 & 1.00 & \textbf{0.78} \\
			\multirow{3}{*}{(1.c)} & $d_1$ & 1.00 & 1.00 & 0.04 & 0.97 & \textbf{0.04} & 1.00 & 1.00 & 0.04 & 0.98 & \textbf{0.04} & 1.00 & 1.00 & 0.10 & 0.99 & \textbf{0.10} & 1.00 & 1.00 & 0.07 & 0.97 & \textbf{0.07} & 1.00 & 1.00 & 0.97 & 0.94 & \textbf{0.92} \\
			& $d_2$ & 1.00 & 1.00 & 0.08 & 0.99 & \textbf{0.07} & 1.00 & 1.00 & 0.08 & 0.99 & \textbf{0.07} & 1.00 & 1.00 & 0.21 & 1.00 & \textbf{0.21} & 1.00 & 1.00 & 0.13 & 0.98 & \textbf{0.13} & 1.00 & 1.00 & 0.99 & 0.97 & \textbf{0.95} \\
			& $d_3$ & 1.00 & 1.00 & 0.11 & 0.99 & \textbf{0.11} & 1.00 & 1.00 & 0.10 & 0.99 & \textbf{0.10} & 1.00 & 1.00 & 0.25 & 1.00 & \textbf{0.25} & 1.00 & 1.00 & 0.19 & 0.99 & \textbf{0.18} & 1.00 & 1.00 & 0.99 & 0.98 & \textbf{0.97} \\
			\bottomrule
		\end{tabular}%
	}
	\vspace{2mm}
	
	\begin{minipage}{0.98\textwidth}
		\footnotesize
		\textbf{Notes:} Numbers stand for proportions over 500 replications. $\mathcal{P}_s$ is the proportion
		that each true active feature is selected. $\mathcal{P}_a$ is the proportion that all true predictors are selected.
	\end{minipage}
\end{table}

\begin{table}[H]
	\centering
	\caption{Quantiles of $\mathcal{S}$ in Euclidean setting.}
	\label{tab:example31-size}
	\resizebox{\textwidth}{!}{%
		\begin{tabular}{l
				*{3}{c}
				*{3}{c}
				*{3}{c}
				*{3}{c}
				*{3}{c}}
			\toprule
			Model
			& \multicolumn{3}{c}{SIS}
			& \multicolumn{3}{c}{SIRS}
			& \multicolumn{3}{c}{DC-SIS}
			& \multicolumn{3}{c}{D-SEVIS}
			& \multicolumn{3}{c}{MDS} \\
			\cmidrule(lr){2-4}\cmidrule(lr){5-7}\cmidrule(lr){8-10}\cmidrule(lr){11-13}\cmidrule(lr){14-16}
			& 25\% & 50\% & 75\%
			& 25\% & 50\% & 75\%
			& 25\% & 50\% & 75\%
			& 25\% & 50\% & 75\%
			& 25\% & 50\% & 75\% \\
			\midrule
			
			\multicolumn{16}{l}{Case 1: $p=2000 \quad \sigma=0.5$ for the AR2 model} \\
			\midrule
			(1.a) & \textbf{4.0} & \textbf{4.0} & \textbf{5.0} & \textbf{4.0} & \textbf{4.0} & \textbf{5.0} & \textbf{4.0} & \textbf{4.0} & \textbf{5.0} & \textbf{4.0} & \textbf{4.0} & \textbf{5.0} & \textbf{4.0} & \textbf{4.0} & \textbf{5.0} \\
			(1.b) & 5.0 & 7.0 & 17.0 & \textbf{4.0} & 7.0 & 19.0 & \textbf{4.0} & \textbf{5.0} & \textbf{8.2} & \textbf{4.0} & 6.0 & 12.0 & \textbf{4.0} & \textbf{5.0} & 10.2 \\
			(1.c) & 526.0 & 1013.5 & 1544.5 & 514.2 & 1008.0 & 1514.5 & 249.8 & 482.5 & 774.8 & 318.2 & 697.5 & 1215.0 & \textbf{4.0} & \textbf{7.5} & \textbf{32.0} \\
			
			\midrule
			\multicolumn{16}{l}{Case 2: $p=2000 \quad \sigma=0.8$ for the AR2 model} \\
			\midrule
			(1.a) & \textbf{7.0} & \textbf{9.0} & \textbf{12.0 } & \textbf{7.0} & \textbf{9.0} & \textbf{12.0} & \textbf{7.0} & \textbf{9.0} & \textbf{12.0} & \textbf{7.0} & \textbf{9.0} & \textbf{12.0} & \textbf{7.0} & \textbf{9.0} & \textbf{12.0} \\
			(1.b) & 43.8 & 152.5 & 489.5 & 44.0 & 184.5 & 577.2 & 19.0 & \textbf{61.0} & \textbf{238.0} & 21.8 & 79.5 & 322.8 & \textbf{16.0} & 67.0 & 261.0 \\
			(1.c) & 126.5 & 502.5 & 1041.2 & 159.5 & 491.5 & 1111.5 & 81.0 & 263.0 & 527.2 & 159.0 & 400.5 & 824.5 & \textbf{6.0} & \textbf{9.0} & \textbf{20.2} \\
			
			\midrule
			\multicolumn{16}{l}{Case 3: $p=5000 \quad \sigma=0.5$ for the AR2 model} \\
			\midrule
			(1.a) & \textbf{4.0} & \textbf{4.0} & \textbf{5.0} & \textbf{4.0} & \textbf{4.0} & \textbf{5.0} & \textbf{4.0} & \textbf{4.0} & \textbf{5.0} & \textbf{4.0} & \textbf{4.0} & \textbf{5.0} & \textbf{4.0} & \textbf{4.0} & \textbf{5.0} \\
			(1.b) & 5.0 & 8.0 & 31.2 & 5.0 & 8.0 & 33.2 & \textbf{4.0} & \textbf{5.0} & \textbf{13.0} & \textbf{4.0} & 6.0 & 20.2 & \textbf{4.0} & \textbf{5.0} & 19.0 \\
			(1.c) & 1230.2 & 2527.0 & 3698.5 & 1275.0 & 2569.0 & 3763.2 & 620.5 & 1222.0 & 1912.0 & 710.5 & 1718.0 & 2903.8 & \textbf{6.0 }& \textbf{16.5} & \textbf{65.0} \\
			
			\midrule
			\multicolumn{16}{l}{Case 4: $p=5000 \quad \sigma=0.8$ for the AR2 model} \\
			\midrule
			(1.a) & \textbf{7.0} & \textbf{9.0} & 11.2 & \textbf{7.0} & \textbf{9.0} & 12.0 & \textbf{7.0} & \textbf{9.0} & 12.0 & \textbf{7.0} & \textbf{9.0} & \textbf{11.0} & \textbf{7.0} & \textbf{9.0} & 12.0 \\
			(1.b) & 63.8 & 331.0 & 1257.0 & 63.8 & 338.0 & 1504.0 & 23.0 & 117.5 & 598.5 & 31.0 & 157.0 & 679.8 & \textbf{20.8} & \textbf{92.0} & \textbf{446.8} \\
			(1.c) & 393.2 & 1188.5 & 2758.2 & 487.5 & 1353.0 & 2868.5 & 262.5 & 692.5 & 1363.5 & 393.0 & 1081.5 & 2410.0 & \textbf{7.0} & \textbf{12.0} & \textbf{43.2} \\
			
			\midrule
			\multicolumn{16}{l}{Case 1: $p=2000 \quad \sigma=0.5$ for the ARMA(1,1) model} \\
			\midrule
			(1.a) & \textbf{4.0} & \textbf{4.0} & \textbf{4.0} & \textbf{4.0} & \textbf{4.0} & \textbf{4.0} & \textbf{4.0} & \textbf{4.0} & \textbf{4.0} & \textbf{4.0} & \textbf{4.0} & \textbf{4.0} & \textbf{4.0} & \textbf{4.0} & \textbf{4.0} \\
			(1.b) & \textbf{4.0} & 5.0 & 6.0 & \textbf{4.0} & 5.0 & 6.0 & \textbf{4.0} & \textbf{4.0} & \textbf{5.0} & \textbf{4.0} & 4.0 & 5.0 & \textbf{4.0} & \textbf{4.0} & \textbf{5.0 }\\
			(1.c) & 482.8 & 1006.0 & 1523.2 & 473.8 & 1032.5 & 1516.0 & 129.0 & 239.0 & 434.0 & 124.8 & 338.5 & 650.0 & \textbf{4.0} & \textbf{4.0 }& \textbf{6.0} \\
			
			\midrule
			\multicolumn{16}{l}{Case 2: $p=2000 \quad \sigma=0.8$ for the ARMA(1,1) model} \\
			\midrule
			(1.a) & \textbf{6.0} & \textbf{8.0} & \textbf{11.0} & 7.0 & \textbf{8.0} & \textbf{11.0} & 7.0 & \textbf{8.0} & \textbf{11.0} & 7.0 & \textbf{8.0} & \textbf{11.0} & 7.0 & \textbf{8.0} & 12.0 \\
			(1.b) & 15.0 & 34.0 & 123.5 & 15.0 & 37.5 & 155.5 & 11.0 & 16.0 & \textbf{39.2} & 12.0 & 20.0 & 66.0 & \textbf{10.0} & \textbf{15.0} & 45.0 \\
			(1.c) & 118.0 & 395.5 & 974.2 & 126.0 & 461.5 & 1037.0 & 46.8 & 131.5 & 279.8 & 58.8 & 176.5 & 455.8 & \textbf{5.0} & \textbf{6.0} & \textbf{8.0} \\
			
			\midrule
			\multicolumn{16}{l}{Case 3: $p=5000 \quad \sigma=0.5$ for the ARMA(1,1) model} \\
			\midrule
			(1.a) & \textbf{4.0} & \textbf{4.0} & \textbf{4.0} & \textbf{4.0} & \textbf{4.0} & \textbf{4.0} & \textbf{4.0} & \textbf{4.0} & \textbf{4.0} & \textbf{4.0} & \textbf{4.0} & \textbf{4.0} & \textbf{4.0} & \textbf{4.0} & 5.0 \\
			(1.b) & \textbf{4.0} & 5.0 & 6.0 & \textbf{4.0} & 5.0 & 6.0 & \textbf{4.0} & \textbf{4.0} & \textbf{5.0} & \textbf{4.0} & \textbf{4.0} & 6.0 & \textbf{4.0} & \textbf{4.0} & 6.0 \\
			(1.c) & 1241.5 & 2489.0 & 3832.2 & 1166.0 & 2472.0 & 3803.5 & 307.8 & 671.0 & 1124.0 & 284.2 & 773.0 & 1717.5 & \textbf{4.0} & \textbf{4.0} & \textbf{9.0} \\
			
			\midrule
			\multicolumn{16}{l}{Case 4: $p=5000 \quad \sigma=0.8$ for the ARMA(1,1) model} \\
			\midrule
			(1.a) & \textbf{6.0} & \textbf{8.0} & \textbf{11.0} & \textbf{6.0} & \textbf{8.0} & \textbf{11.0} & 7.0 & \textbf{8.0} & \textbf{11.0} & 7.0 & \textbf{8.0} & \textbf{11.0} & 7.0 & \textbf{8.0} & \textbf{11.0} \\
			(1.b) & 17.0 & 54.0 & 200.5 & 19.0 & 68.0 & 256.2 & 12.0 & \textbf{20.0} & \textbf{62.0} & 12.8 & 27.0 & 122.5 & \textbf{10.0} & 22.0 & 87.0 \\
			(1.c) & 317.5 & 940.0 & 2588.2 & 370.8 & 1136.0 & 2773.8 & 105.2 & 316.5 & 664.8 & 157.0 & 476.0 & 1160.5 & \textbf{5.0} & \textbf{6.0} & \textbf{10.0} \\
			
			\bottomrule
		\end{tabular}%
	}
	
	\vspace{2mm}
	\begin{minipage}{0.98\textwidth}
		\footnotesize
		\textbf{Notes:} $\mathcal{S}$ is the size of the minimum set to contain all the active features.
		Its 25\%, 50\%, and 75\% quantiles are reported in the tables.
	\end{minipage}
\end{table}

Tables~\ref{tab:ar2-euclidean}--\ref{tab:example31-size} show that the performance of MDS varies across the three models, but is overall competitive with the benchmark methods. In model (1.a), all methods perform very well, and MDS is slightly below the best Euclidean competitors in a few configurations. However, the differences are small, and its overall screening performance remains comparable. In model (1.b), the performance of MDS becomes more favorable. From the tables, its $\mathcal P_s$ and $\mathcal P_a$ values are generally close to those of the best competing methods, and the corresponding quantiles of $\mathcal S$ are also at a similar level. In model (1.c), the advantage of MDS becomes much more evident. The tables show that the Euclidean benchmark methods often have very low selection frequencies for the nonlinear active feature $X_{12}$ and usually require much larger screened set sizes to recover all active predictors, whereas MDS attains much higher recovery probabilities and substantially smaller values of $\mathcal S$.

These conclusions are similar under both the AR(2) and ARMA(1,1) designs. Taken together, the Euclidean simulation results show that MDS remains competitive when the signal is linear and performs much better when the active structure is more nonlinear.

\subsection{Non-Euclidean setting}\label{subsec:sim-noneuclid}
We next consider a setting in which the predictor $X_i\in\mathbb R^p$ is generated in the same way as in the Euclidean setting, while the response $Y_i$ is object valued and lies in a non-Euclidean metric space. Since the Euclidean competitors, namely SIS, SIRS, DC-SIS, and D-SEVIS, rely on Euclidean vector structure and are not designed for object valued targets, we focus here on the finite sample performance of MDS.

Throughout this subsection, we consider $p\in\{2000,5000\}$ and use the same time series designs for $\{X_i\}$ as in Section~\ref{subsec:sim-euclid}. We take the cross sectional correlation parameter to be $\sigma\in\{0,0.5\}$.

We consider a Wasserstein response for which $Y_i$ is a univariate distribution represented by its quantile function, so that $Y_i$ is viewed as an element of the quadratic Wasserstein space $\mathcal W_2(\mathbb R)$. The distance is measured by the usual $W_2$ metric,
\[
d_W^2(F,G)=\int_0^1\bigl(F^{-1}(u)-G^{-1}(u)\bigr)^2\,du,
\]
where $F^{-1}$ and $G^{-1}$ are the corresponding quantile functions. Let $\beta\in\mathbb R^p$ be sparse with active coordinates 
\[
\mathcal A^\star=\{1,\,2,\,12,\,22\},\quad 
\beta_j=1\ \text{for }j\in\mathcal A^\star,\ \text{and }\beta_j=0\ \text{for }j\notin\mathcal A^\star.
\]
Define the scalar index $\eta_i:=X_i^\top\beta$. The response is generated by
\[
Q_{Y_i}(u)=\eta_i+0.1\,\Phi^{-1}(u),\qquad u\in(0,1),
\]
where $\Phi^{-1}$ is the standard normal quantile function.

In implementation, we evaluate $Q_{Y_i}(u)$ on a fine grid of $u\in(0,1)$ and treat the resulting curve as the observed object for $Y_i$. Under the $W_2$ metric, this is equivalent to an $L^2$ distance between quantile curves. We then apply MDS to screen predictors using the metric dependence score between $X^k$ and the response object.

\begin{table}[!htbp]
	\centering
	\caption{The proportions of $\mathcal{P}_s$ and $\mathcal{P}_a$ in the non-Euclidean setting for MDS.}
	\label{tab:mds-pa-non-euclidean}
	\renewcommand{\arraystretch}{0.85}
	\resizebox{0.9\textwidth}{!}{%
		\begin{tabular}{ccc ccccc ccccc}
			\toprule
			\multicolumn{3}{c}{} & \multicolumn{5}{c}{AR2} & \multicolumn{5}{c}{ARMA(1,1)} \\
			\cmidrule(lr){4-8}\cmidrule(lr){9-13}
			\multicolumn{3}{c}{}
			& \multicolumn{4}{c}{$\mathcal{P}_s$} & $\mathcal{P}_a$
			& \multicolumn{4}{c}{$\mathcal{P}_s$} & $\mathcal{P}_a$ \\
			\cmidrule(lr){4-7}\cmidrule(lr){9-12}
			$p$ & $\sigma$ & $d$
			& $X_1$ & $X_2$ & $X_{12}$ & $X_{22}$ & ALL
			& $X_1$ & $X_2$ & $X_{12}$ & $X_{22}$ & ALL \\
			\midrule
			\multirow{3}{*}{2000} & \multirow{3}{*}{0}
			& $d_1$ & 1.00 & 1.00 & 1.00 & 1.00 & \textbf{1.00}
			& 1.00 & 1.00 & 1.00 & 1.00 & \textbf{1.00} \\
			& & $d_2$ & 1.00 & 1.00 & 1.00 & 1.00 & \textbf{1.00}
			& 1.00 & 1.00 & 1.00 & 1.00 & \textbf{1.00} \\
			& &  $d_3$ & 1.00 & 1.00 & 1.00 & 1.00 & \textbf{1.00}
			& 1.00 & 1.00 & 1.00 & 1.00 & \textbf{1.00} \\
			\midrule
			\multirow{3}{*}{2000} & \multirow{3}{*}{0.5}
			& $d_1$ & 1.00 & 1.00 & 1.00 & 0.99 & \textbf{0.98}
			& 1.00 & 1.00 & 1.00 & 1.00 & \textbf{1.00} \\
			& &  $d_2$ & 1.00 & 1.00 & 1.00 & 0.99 & \textbf{0.99}
			& 1.00 & 1.00 & 1.00 & 1.00 & \textbf{1.00} \\
			& &  $d_3$ & 1.00 & 1.00 & 1.00 & 0.99 & \textbf{0.99}
			& 1.00 & 1.00 & 1.00 & 1.00 & \textbf{1.00} \\
			\midrule
			\multirow{3}{*}{5000} & \multirow{3}{*}{0}
			& $d_1$ & 1.00 & 1.00 & 0.99 & 1.00 & \textbf{0.99}
			& 1.00 & 1.00 & 1.00 & 1.00 & \textbf{1.00} \\
			& &  $d_2$ & 1.00 & 1.00 & 1.00 & 1.00 & \textbf{0.99}
			& 1.00 & 1.00 & 1.00 & 1.00 & \textbf{1.00} \\
			& &  $d_3$ & 1.00 & 1.00 & 1.00 & 1.00 & \textbf{0.99}
			& 1.00 & 1.00 & 1.00 & 1.00 & \textbf{1.00} \\
			\midrule
			\multirow{3}{*}{5000} & \multirow{3}{*}{0.5}
			& $d_1$ & 1.00 & 1.00 & 0.97 & 0.96 & \textbf{0.94}
			& 1.00 & 1.00 & 1.00 & 0.99 & \textbf{0.99} \\
			& &  $d_2$ & 1.00 & 1.00 & 0.99 & 0.98 & \textbf{0.97}
			& 1.00 & 1.00 & 1.00 & 0.99 & \textbf{0.99} \\
			& &  $d_3$ & 1.00 & 1.00 & 0.99 & 0.98 & \textbf{0.98}
			& 1.00 & 1.00 & 1.00 & 1.00 & \textbf{1.00} \\
			\bottomrule
		\end{tabular}%
	}
	
	\vspace{2mm}
	\begin{minipage}{0.95\textwidth}
		\footnotesize
		\textbf{Notes:} Numbers stand for proportions over 500 replications. $\mathcal{P}_s$ is the proportion that each true active feature is selected, and $\mathcal{P}_a$ is the proportion that all true predictors are selected.
	\end{minipage}
\end{table}

\begin{table}[!htbp]
	\centering
	\caption{The quantiles of $\mathcal{S}$ in the non-Euclidean setting for MDS.}
	\label{tab:mds-size-non-euclidean}
	\renewcommand{\arraystretch}{0.85}
	\resizebox{0.66\textwidth}{!}{%
		\begin{tabular}{cc ccc ccc}
			\toprule
			\multicolumn{2}{c}{} & \multicolumn{3}{c}{AR2} & \multicolumn{3}{c}{ARMA(1,1)} \\
			\cmidrule(lr){3-5}\cmidrule(lr){6-8}
			$p$ & $\sigma$ & 25\% & 50\% & 75\% & 25\% & 50\% & 75\% \\
			\midrule
			2000 & 0   & 4.0 & 4.0 & 4.0 & 4.0 & 4.0 & 4.0 \\
			2000 & 0.5 & 4.0 & 4.0 & 5.0 & 4.0 & 4.0 & 4.0 \\
			5000 & 0   & 4.0 & 4.0 & 4.0 & 4.0 & 4.0 & 4.0 \\
			5000 & 0.5 & 4.0 & 4.0 & 6.2 & 4.0 & 4.0 & 5.0 \\
			\bottomrule
		\end{tabular}%
	}
	
	\vspace{2mm}
	\begin{minipage}{0.9\textwidth}
		\footnotesize
		\textbf{Notes:} $\mathcal{S}$ is the size of the minimum set to contain all the active features.
		Its 25\%, 50\%, and 75\% quantiles are reported in the table.
	\end{minipage}
\end{table}

Tables~\ref{tab:mds-pa-non-euclidean}--\ref{tab:mds-size-non-euclidean} report the results in the Wasserstein setting. From the tables, MDS achieves very high selection frequencies for all active predictors across all configurations, and $\mathcal P_a$ remains close to one even when $p$ increases and $\sigma>0$. The quantiles of $\mathcal S$ are also small. In most cases, they are very close to the oracle lower bound $|\mathcal A^\star|=4$. This shows that only a small screened set is typically needed to recover all active predictors. These results indicate that MDS remains effective for object valued responses in a representative non-Euclidean metric space under serial dependence and ultrahigh dimensionality.

These simulation results show that MDS can handle both Euclidean and non-Euclidean responses effectively. It remains competitive with existing methods in the Euclidean setting and continues to achieve high recovery probabilities with small screened set sizes in the non-Euclidean setting.

\section{Empirical Study}\label{sec:empirical}
This section evaluates the MDS procedure in a large universe asset selection problem using the Chinese A-share market. The investment universe contains all A shares listed on the Shanghai and Shenzhen exchanges from 2023-07-01 to 2025-12-31. We use 5 minute intraday prices for the target construction and one minute intraday returns to estimate the intraday spot volatility curve. We retain stocks with complete observations over the sample period. After this filtering, the final universe contains $p=2938$ stocks.

We implement a monthly rebalancing backtest with a rolling estimation window of $M=6$ months and a one month holding period. At each rebalancing date, we use the preceding six months of data to construct the daily series $\{Y_t\}$ and the observed covariate series $\{\widehat X_{i,t}\}$ defined in Section~\ref{subsec:setup}. For each stock, we compute a marginal relevance score with respect to the target, rank all stocks by this score, and select the top $d$ names, where $d_1=30$, $d_2=60$, and $d_3=90$. The selected portfolio is then held for one month. After that, the window rolls forward by one month and the procedure is repeated. We report results without transaction costs. Over the sample period, this yields 24 out of sample portfolio observations. We refer to the corresponding sequence of monthly portfolio outcomes as the holding period performance series.

The MDS score is estimated using a partition of the target space as described in Section~\ref{subsubsec:MDS-estimation}. In each rolling window, we partition the realized target values $\{Y_t\}$ into $H=8$ slices using empirical quantiles, which yields sets $\{\Omega_h\}_{h=1}^H$ with approximately equal empirical probabilities. These slices define the within slice Fr\'echet means and variances used in the plug-in MDS estimator for ranking.

In each rolling window, we construct the observed point curve covariates $\widehat X_{i,t}=(R_{i,t},\widehat v_{i,t})$, where $\widehat v_{i,t}$ is the empirical grid estimator of the latent intraday spot volatility curve $v_{i,t}=\sigma_{i,t}$. We then compute pairwise distances using the empirical version of $d_X$ described in Section~\ref{subsubsec:objects-X}. We keep the same distance specification across all rolling windows and across all methods in the empirical analysis.

We compare MDS based selection with two additional selection rules that are standard in this screening first portfolio pipeline: momentum based ranking (MTM) and the dependence screening score in \citet{wang2022asset} (D-SEVIS). In each rolling window, all three methods produce a ranked list of the $p$ stocks, and we form the candidate set by keeping the top $d\in\{30,60,90\}$ names. Table~\ref{tab:emp-methods} summarizes the three selection rules and the weighting schemes described below.
\begin{table}[htbp]
	\centering
	\caption{Selection and weighting methods in the empirical study.}
	\label{tab:emp-methods}
	\renewcommand{\arraystretch}{1.0}
	\resizebox{\linewidth}{!}{%
		\begin{tabular}{ll}
			\toprule
			\textbf{Stock selection method} & \textbf{Symbol} \\
			\midrule
			Metric Dependence Screening & MDS \\
			Dependence screening score of \citet{wang2022asset} & D-SEVIS \\
			Momentum strategies & MTM \\
			\midrule
			\textbf{Weighting method} & \textbf{Symbol} \\
			\midrule
			Equal weight & EW \\
			Score-squared weight & SEVW \\
			Global optimization combination & GOC \\
			Global optimization combination with shrinkage $\Sigma$ in \citet{schafer2005shrinkage} & GOCS \\
			\bottomrule
		\end{tabular}%
	}
\end{table}

Given a selected set of size $d$, we form portfolios under four long only weighting schemes. For the optimization based schemes GOC and GOCS, the weights are obtained by solving the following constrained global minimum variance problem on the selected stocks:
\[
\min_{\omega\in\mathbb R^{d}}\ \omega^\top \widehat\Sigma_t\,\omega
\quad\text{s.t.}\quad
\mathbf 1^\top \omega=1,\qquad 0\le \omega_j\le u_d,\ \ j=1,\ldots,d,
\]
where the long only constraint reflects the A-share short sale restriction and the cap $u_d$ controls concentration. We set
\[
u_{30}=0.15,\qquad u_{60}=0.10,\qquad u_{90}=0.05.
\]
The two schemes differ only in the covariance input. GOC uses the sample covariance matrix, whereas GOCS uses a shrinkage covariance estimator.

\begin{figure}[H]
	\centering
	\includegraphics[width=1\linewidth]{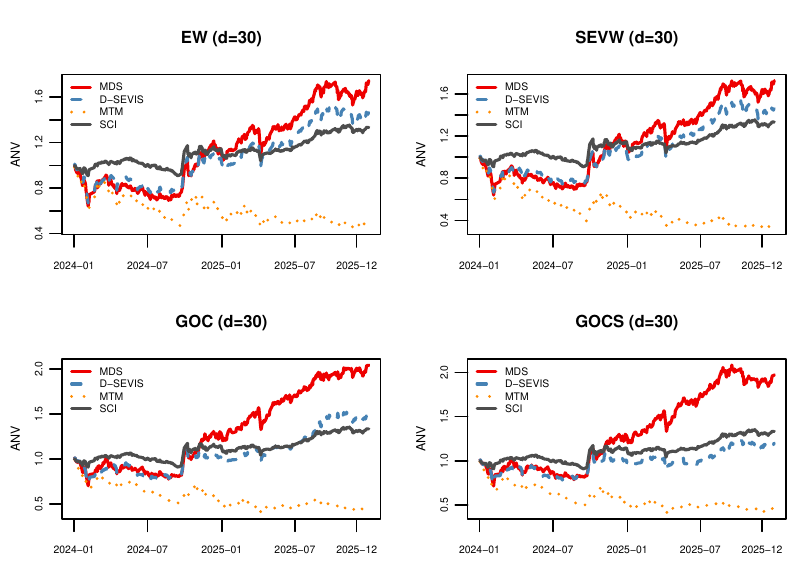}
	\caption{Accumulated net value paths for $d=30$ under four weighting schemes (EW, SEVW, GOC, and GOCS) in 2024--2025.}
	\label{fig:emp-nav-d30}
\end{figure}

Portfolio performance is evaluated using the monthly holding period return series generated by the rolling backtest. The results are reported in Table~\ref{tab:emp-results-2024-2025} and Figure~\ref{fig:emp-nav-d30}. We report accumulated net value, maximum drawdown, and Sharpe ratio for each combination of selection rule, weighting scheme, and selection size $d\in\{30,60,90\}$. The Sharpe ratio is annualized with an annualization factor of $252$ and an annual risk free rate of $0.04$. For comparison, we also report the same performance metrics for the Shanghai Composite Index (SCI).

Table~\ref{tab:emp-results-2024-2025} shows that the MDS portfolios outperform the two benchmarks over 2024--2025. First, across all weighting schemes and all three selection sizes, MDS attains higher accumulated net values than both D-SEVIS and MTM. Its strongest performance appears under the optimization based schemes GOC and GOCS. The best result is attained by MDS with GOC at $d=30$, which reaches an accumulated net value of $2.04$ and a Sharpe ratio of $1.22$. Second, these gains are accompanied by better risk adjusted performance. Sharpe ratios under MDS are positive in all settings and are clearly higher than those of the competing methods. The difference is especially clear under GOC and GOCS, where the Sharpe ratios exceed 1 for $d=30$ and $d=60$. By contrast, the momentum benchmark MTM performs poorly in this sample, with accumulated net values below $1$ and negative Sharpe ratios under every weighting scheme. Third, the drawdown results suggest that MDS does not obtain its gains by taking materially larger downside risk. Under EW and SEVW, the maximum drawdowns of MDS are comparable to those of D-SEVIS, while under GOC and GOCS they are generally comparable or smaller in magnitude. For example, at $d=30$, the MDS portfolio under GOCS has the same maximum drawdown as its D-SEVIS counterpart but a substantially higher Sharpe ratio.

The relative performance of the weighting schemes within each selection rule is also informative. For both MDS and D-SEVIS, the score based weighting scheme SEVW performs similarly to equal weighting EW, whereas the optimization based schemes, especially GOCS, usually deliver higher Sharpe ratios. This pattern is consistent with the role of the long only cap constraints in controlling concentration while allowing covariance aware reallocation among the selected stocks. More importantly, MDS remains the best performing selection rule in terms of both accumulated net value and Sharpe ratio under every weighting scheme, which suggests that its advantage comes mainly from the screening stage rather than from a particular weighting rule.

\begin{table}[!htbp]
	\centering
	\caption{Performance of MDS, D-SEVIS and MTM portfolios in 2024--2025.}
	\label{tab:emp-results-2024-2025}
	\renewcommand{\arraystretch}{0.8}
	\begin{threeparttable}
		\resizebox{\linewidth}{!}{%
			\begin{tabular}{clcccc}
				\toprule
				Selection & Weighting & \# stocks & Accumulated net value & Max drawdown & Sharpe ratio \\
				\midrule
				\multirow{12}{*}{MDS}
				& EW   & 30 & 1.74 & -0.36 & 0.90 \\
				& EW   & 60 & 1.69 & -0.35 & 0.86 \\
				& EW   & 90 & 1.61 & -0.35 & 0.79 \\
				& SEVW & 30 & 1.72 & -0.36 & 0.89 \\
				& SEVW & 60 & 1.69 & -0.35 & 0.86 \\
				& SEVW & 90 & 1.62 & -0.35 & 0.79 \\
				& GOC  & 30 & 2.04 & -0.31 & 1.22 \\
				& GOC  & 60 & 1.84 & -0.27 & 1.10 \\
				& GOC  & 90 & 1.68 & -0.27 & 0.96 \\
				& GOCS & 30 & 1.98 & -0.29 & 1.17 \\
				& GOCS & 60 & 1.76 & -0.25 & 1.02 \\
				& GOCS & 90 & 1.65 & -0.27 & 0.92 \\
				\midrule
				\multirow{12}{*}{D-SEVIS}
				& EW   & 30 & 1.46 & -0.34 & 0.65 \\
				& EW   & 60 & 1.45 & -0.34 & 0.64 \\
				& EW   & 90 & 1.48 & -0.34 & 0.67 \\
				& SEVW & 30 & 1.46 & -0.34 & 0.64 \\
				& SEVW & 60 & 1.45 & -0.34 & 0.64 \\
				& SEVW & 90 & 1.47 & -0.34 & 0.66 \\
				& GOC  & 30 & 1.47 & -0.31 & 0.70 \\
				& GOC  & 60 & 1.38 & -0.28 & 0.60 \\
				& GOC  & 90 & 1.39 & -0.28 & 0.62 \\
				& GOCS & 30 & 1.20 & -0.29 & 0.34 \\
				& GOCS & 60 & 1.22 & -0.27 & 0.36 \\
				& GOCS & 90 & 1.26 & -0.27 & 0.43 \\
				\midrule
				\multirow{12}{*}{MTM}
				& EW   & 30 & 0.51 & -0.55 & -0.67 \\
				& EW   & 60 & 0.64 & -0.45 & -0.50 \\
				& EW   & 90 & 0.76 & -0.40 & -0.30 \\
				& SEVW & 30 & 0.37 & -0.67 & -0.95 \\
				& SEVW & 60 & 0.43 & -0.62 & -0.87 \\
				& SEVW & 90 & 0.48 & -0.57 & -0.79 \\
				& GOC  & 30 & 0.45 & -0.60 & -1.08 \\
				& GOC  & 60 & 0.70 & -0.41 & -0.60 \\
				& GOC  & 90 & 0.88 & -0.29 & -0.24 \\
				& GOCS & 30 & 0.45 & -0.59 & -1.05 \\
				& GOCS & 60 & 0.65 & -0.42 & -0.71 \\
				& GOCS & 90 & 0.86 & -0.30 & -0.28 \\
				\midrule
				\multirow{1}{*}{SCI\tnote{a}} & -- & -- & 1.33 & -0.15 & 0.73 \\
				\bottomrule
			\end{tabular}%
		}
		\begin{tablenotes}[flushleft]
			\footnotesize
			\item[a] SCI denotes the Shanghai Composite Index.
		\end{tablenotes}
	\end{threeparttable}
\end{table}

Figure~\ref{fig:emp-nav-d30} plots the accumulated net value paths for $d=30$ under the four weighting schemes and provides a time series view of the results in Table~\ref{tab:emp-results-2024-2025}. In the early part of the evaluation period, all strategies experience an initial decline, which coincides with a broad market drawdown in the A-share market. As market conditions improve, the screening based strategies recover and their performance paths begin to separate. Across all four weighting schemes, the MDS portfolio experiences an early drawdown and may lag the SCI at the beginning of the evaluation period, but it then recovers, overtakes the index, and remains ahead for most of 2025, with the gap widening toward the end of the sample. The MDS trajectory also stays above the D-SEVIS and MTM trajectories for most of the period, which suggests that the outperformance is cumulative rather than driven by a single month. The four panels further show that the optimization based weighting schemes, GOC and GOCS, amplify the gains from screening. Relative to EW and SEVW, they exhibit a faster and more persistent rise in accumulated net value after the early drawdown. This pattern is consistent with the additional diversification and risk control provided by the constrained minimum variance allocation on the screened set.

Taken together, Table~\ref{tab:emp-results-2024-2025} and Figure~\ref{fig:emp-nav-d30} show that the proposed covariate construction and metric based screening are empirically effective. By incorporating an estimated intraday spot volatility curve into the asset covariate $\widehat X_{i,t}$, MDS captures risk state variation that is useful for ranking assets against the risk adjusted target. This leads to economically meaningful improvements in out of sample portfolio performance under multiple portfolio construction rules.

\section{Conclusion}
\label{sec:Conclusion}

This paper proposes MDS for large universe asset selection with object valued covariate information. Each asset day observation is represented by a point curve object that combines the daily return with a latent intraday spot volatility curve. Under a continuous time diffusion representation of the within day log price process, this curve is defined as $v(t)=\sigma(t)$ and is estimated from high frequency intraday returns in the empirical implementation. Assets are then ranked by a metric dependence score relative to a risk adjusted target. The resulting dependence coefficient is defined through Fr\'echet means and variance and admits an explained variation interpretation. This keeps it close in spirit to mean-variance analysis while allowing it to operate in general metric spaces. We develop a feasible estimator based on slicing the target space and establish screening guarantees for dependent time series in ultrahigh dimensions. In the empirical study based on a large A share universe with high frequency intraday data, MDS delivers economically meaningful portfolio gains. Across selection sizes and long only weighting schemes, it improves accumulated net values and Sharpe ratios relative to D-SEVIS and MTM while keeping drawdowns broadly comparable.

Several directions remain for future study. The framework is not limited to the intraday spot volatility curve used here. It can incorporate alternative high frequency risk state objects, including jump robust volatility measures and multi resolution intraday features. It is also important to extend the empirical pipeline to account for turnover, transaction costs, and liquidity constraints. Finally, multivariate or conditional targets and time varying portfolio constraints may improve robustness to regime changes and extend the framework to more general asset allocation problems.








\newpage
\bibliographystyle{elsarticle-harv}
\bibliography{references}

@article{11475778,
	author={Qiu, Rui and Yao, Fang and Yu, Zhou},
	journal={IEEE Transactions on Information Theory}, 
	title={Fréchet Regression with Mondrian Forests: Finite-Sample Guarantees and Ensemble Benefits}, 
	year={2026},
	volume={},
	number={},
	pages={1-1}}

@article{ait2008out,
	title={Out of sample forecasts of quadratic variation},
	author={A{\"\i}t-Sahalia, Yacine and Mancini, Loriano},
	journal={Journal of Econometrics},
	volume={147},
	number={1},
	pages={17--33},
	year={2008},
	publisher={Elsevier}
}

@article{andersen1997intraday,
	title={Intraday periodicity and volatility persistence in financial markets},
	author={Andersen, Torben G and Bollerslev, Tim},
	journal={Journal of empirical finance},
	volume={4},
	number={2-3},
	pages={115--158},
	year={1997},
	publisher={Elsevier}
}

@article{andersen2001distribution,
	title={The distribution of realized stock return volatility},
	author={Andersen, Torben G and Bollerslev, Tim and Diebold, Francis X and Ebens, Heiko},
	journal={Journal of financial economics},
	volume={61},
	number={1},
	pages={43--76},
	year={2001},
	publisher={Elsevier}
}

@article{andersen2003modeling,
	title={Modeling and forecasting realized volatility},
	author={Andersen, Torben G and Bollerslev, Tim and Diebold, Francis X and Labys, Paul},
	journal={Econometrica},
	volume={71},
	number={2},
	pages={579--625},
	year={2003},
	publisher={Wiley Online Library}
}

@article{barndorff2002econometric,
	title={Econometric analysis of realized volatility and its use in estimating stochastic volatility models},
	author={Barndorff-Nielsen, Ole E and Shephard, Neil},
	journal={Journal of the Royal Statistical Society Series B: Statistical Methodology},
	volume={64},
	number={2},
	pages={253--280},
	year={2002},
	publisher={Oxford University Press}
}

@article{barndorff2004power,
	title={Power and bipower variation with stochastic volatility and jumps},
	author={Barndorff-Nielsen, Ole E and Shephard, Neil},
	journal={Journal of financial econometrics},
	volume={2},
	number={1},
	pages={1--37},
	year={2004},
	publisher={Oxford University Press}
}

@article{best1991sensitivity,
	title={On the sensitivity of mean-variance-efficient portfolios to changes in asset means: some analytical and computational results},
	author={Best, Michael J and Grauer, Robert R},
	journal={The review of financial studies},
	volume={4},
	number={2},
	pages={315--342},
	year={1991},
	publisher={Oxford University Press}
}

@article{britten1999sampling,
	title={The sampling error in estimates of mean-variance efficient portfolio weights},
	author={Britten-Jones, Mark},
	journal={The Journal of Finance},
	volume={54},
	number={2},
	pages={655--671},
	year={1999},
	publisher={Wiley Online Library}
}

@article{bhattacharjee2023single,
	title={Single index Fr{\'e}chet regression},
	author={Bhattacharjee, Satarupa and M{\"u}ller, Hans-Georg},
	journal={The Annals of Statistics},
	volume={51},
	number={4},
	pages={1770},
	year={2023},
	publisher={Institute of Mathematical Statistics}
}

@article{bollerslev2000intraday,
	title={Intraday periodicity, long memory volatility, and macroeconomic announcement effects in the US Treasury bond market},
	author={Bollerslev, Tim and Cai, Jun and Song, Frank M},
	journal={Journal of empirical finance},
	volume={7},
	number={1},
	pages={37--55},
	year={2000},
	publisher={Elsevier}
}

@book{campbell2002strategic,
	title={Strategic asset allocation: portfolio choice for long-term investors},
	author={Campbell, John Y and Viceira, Luis M},
	year={2002},
	publisher={Clarendon Lectures in Economic}
}

@article{chen2017sure,
	title={Sure explained variability and independence screening},
	author={Chen, Min and Lian, Yimin and Chen, Zhao and Zhang, Zhengjun},
	journal={Journal of Nonparametric Statistics},
	volume={29},
	number={4},
	pages={849--883},
	year={2017},
	publisher={Taylor \& Francis}
}

@article{chopra1993effect,
	title={The effect of errors in means, variances, and covariances on optimal portfolio choice},
	author={Chopra, Vijay K and Ziemba, William T and others},
	journal={Journal of Portfolio Management},
	volume={19},
	number={2},
	pages={6--11},
	year={1993},
	publisher={World Scientific}
}

@article{demiguel2009optimal,
	title={Optimal versus naive diversification: How inefficient is the 1/N portfolio strategy?},
	author={DeMiguel, Victor and Garlappi, Lorenzo and Uppal, Raman},
	journal={The review of Financial studies},
	volume={22},
	number={5},
	pages={1915--1953},
	year={2009},
	publisher={Oxford University Press}
}

@article{fan2008sure,
	title={Sure independence screening for ultrahigh dimensional feature space},
	author={Fan, Jianqing and Lv, Jinchi},
	journal={Journal of the Royal Statistical Society Series B: Statistical Methodology},
	volume={70},
	number={5},
	pages={849--911},
	year={2008},
	publisher={Oxford University Press}
}

@article{fan2010selective,
	title={A selective overview of variable selection in high dimensional feature space},
	author={Fan, Jianqing and Lv, Jinchi},
	journal={Statistica Sinica},
	volume={20},
	number={1},
	pages={101},
	year={2010}
}

@article{fan2012vast,
	title={Vast portfolio selection with gross-exposure constraints},
	author={Fan, Jianqing and Zhang, Jingjin and Yu, Ke},
	journal={Journal of the American Statistical Association},
	volume={107},
	number={498},
	pages={592--606},
	year={2012},
	publisher={Taylor \& Francis}
}

@article{fleming2001economic,
	title={The economic value of volatility timing},
	author={Fleming, Jeff and Kirby, Chris and Ostdiek, Barbara},
	journal={The Journal of Finance},
	volume={56},
	number={1},
	pages={329--352},
	year={2001},
	publisher={Wiley Online Library}
}

@article{frost1986empirical,
	title={An empirical Bayes approach to efficient portfolio selection},
	author={Frost, Peter A and Savarino, James E},
	journal={Journal of Financial and Quantitative Analysis},
	volume={21},
	number={3},
	pages={293--305},
	year={1986},
	publisher={Cambridge University Press}
}

@article{garlappi2007portfolio,
	title={Portfolio selection with parameter and model uncertainty: A multi-prior approach},
	author={Garlappi, Lorenzo and Uppal, Raman and Wang, Tan},
	journal={The Review of Financial Studies},
	volume={20},
	number={1},
	pages={41--81},
	year={2007},
	publisher={Oxford University Press}
}

@article{green1992will,
	title={When will mean-variance efficient portfolios be well diversified?},
	author={Green, Richard C and Hollifield, Burton},
	journal={The Journal of Finance},
	volume={47},
	number={5},
	pages={1785--1809},
	year={1992},
	publisher={Wiley Online Library}
}

@inproceedings{gretton2005measuring,
	title={Measuring statistical dependence with Hilbert-Schmidt norms},
	author={Gretton, Arthur and Bousquet, Olivier and Smola, Alex and Sch{\"o}lkopf, Bernhard},
	booktitle={International conference on algorithmic learning theory},
	pages={63--77},
	year={2005},
	organization={Springer}
}

@article{gretton2007kernel,
	title={A kernel statistical test of independence},
	author={Gretton, Arthur and Fukumizu, Kenji and Teo, Choon and Song, Le and Sch{\"o}lkopf, Bernhard and Smola, Alex},
	journal={Advances in neural information processing systems},
	volume={20},
	year={2007}
}

@misc{he2026frechetcorrelationcoefficientheterogeneous,
  title={The Fr{\'e}chet correlation coefficient for heterogeneous random objects},
  author={Shuaida He and Yangzhou Chen and Xin Chen},
  year={2026},
  note={arXiv preprint arXiv:2604.10482}
}

@article{jagannathan2003risk,
	title={Risk reduction in large portfolios: Why imposing the wrong constraints helps},
	author={Jagannathan, Ravi and Ma, Tongshu},
	journal={The journal of finance},
	volume={58},
	number={4},
	pages={1651--1683},
	year={2003},
	publisher={Wiley Online Library}
}

@article{jobson1980estimation,
	title={Estimation for Markowitz efficient portfolios},
	author={Jobson, J David and Korkie, Bob},
	journal={Journal of the American Statistical Association},
	volume={75},
	number={371},
	pages={544--554},
	year={1980},
	publisher={Taylor \& Francis}
}

@article{jorion1986bayes,
	title={Bayes-Stein estimation for portfolio analysis},
	author={Jorion, Philippe},
	journal={Journal of Financial and Quantitative analysis},
	volume={21},
	number={3},
	pages={279--292},
	year={1986},
	publisher={Cambridge University Press}
}

@article{kan2007optimal,
	title={Optimal portfolio choice with parameter uncertainty},
	author={Kan, Raymond and Zhou, Guofu},
	journal={Journal of Financial and Quantitative Analysis},
	volume={42},
	number={3},
	pages={621--656},
	year={2007},
	publisher={Cambridge University Press}
}

@article{ledoit2003honey,
	title={Honey, I Shrunk the Sample Covariance Matrix},
	author={Ledoit, Olivier and Wolf, Michael},
	journal={The Journal of Portfolio Management},
	volume={30},
	number={4},
	pages={110--119},
	year={2004},
	publisher={Portfolio Management Research}
}

@article{ledoit2003improved,
	title={Improved estimation of the covariance matrix of stock returns with an application to portfolio selection},
	author={Ledoit, Olivier and Wolf, Michael},
	journal={Journal of empirical finance},
	volume={10},
	number={5},
	pages={603--621},
	year={2003},
	publisher={Elsevier}
}

@article{li2012feature,
	title={Feature screening via distance correlation learning},
	author={Li, Runze and Zhong, Wei and Zhu, Liping},
	journal={Journal of the American Statistical Association},
	volume={107},
	number={499},
	pages={1129--1139},
	year={2012},
	publisher={Taylor \& Francis}
}

@article{lyons2013distance,
	title={Distance covariance in metric spaces},
	author={Lyons, Russell},
	journal={The Annals of Probability},
	pages={3284--3305},
	year={2013},
	publisher={JSTOR}
}

@article{markowitz1952,
	author = {Markowitz, Harry},
	title = {PORTFOLIO SELECTION},
	journal = {The Journal of Finance},
	volume = {7},
	number = {1},
	pages = {77-91},
	year = {1952}
}

@book{markowitz2008portfolio,
	title={Portfolio selection: efficient diversification of investments},
	author={Markowitz, Harry M},
	year={2008},
	publisher={Yale university press}
}

@article{martens2007measuring,
	title={Measuring volatility with the realized range},
	author={Martens, Martin and Van Dijk, Dick},
	journal={Journal of Econometrics},
	volume={138},
	number={1},
	pages={181--207},
	year={2007},
	publisher={Elsevier}
}

@article{michaud1989markowitz,
	title={The Markowitz optimization enigma: Is ‘optimized’optimal?},
	author={Michaud, Richard O},
	journal={Financial analysts journal},
	volume={45},
	number={1},
	pages={31--42},
	year={1989},
	publisher={Taylor \& Francis}
}

@article{pan2020ball,
	title={Ball covariance: A generic measure of dependence in banach space},
	author={Pan, Wenliang and Wang, Xueqin and Zhang, Heping and Zhu, Hongtu and Zhu, Jin},
	journal={Journal of the American Statistical Association},
	year={2020},
	publisher={Taylor \& Francis}
}

@article{petersen2019frechet,
	title={Fr{\'e}chet regression for random objects with Euclidean predictors},
	author={Petersen, Alexander and M{\"u}ller, Hans-Georg},
	journal={The Annals of Statistics},
	volume={47},
	number={2},
	pages={691},
	year={2019},
	publisher={Institute of Mathematical Statistics}
}

@article{schafer2005shrinkage,
	title={A shrinkage approach to large-scale covariance matrix estimation and implications for functional genomics},
	author={Sch{\"a}fer, Juliane and Strimmer, Korbinian},
	journal={Statistical applications in genetics and molecular biology},
	volume={4},
	number={1},
	year={2005},
	publisher={De Gruyter}
}

@article{szekely2007measuring,
	title={Measuring and testing dependence by correlation of distances},
	author={Sz{\'e}kely, GJ and Rizzo, ML and Bakirov, NK},
	journal={Annals of Statistics},
	volume={35},
	number={6},
	pages={2769--2794},
	year={2007},
	publisher={Institute of Mathematical Statistics}
}

@article{szekely2009brownian,
	title={Brownian distance covariance},
	author={Sz{\'e}kely, G{\'a}bor J and Rizzo, Maria L},
	journal={The Annals of Applied Statistics},
	volume={3},
	number={4},
	pages={1236},
	year={2009},
	publisher={Institute of Mathematical Statistics}
}

@article{wang2022asset,
	title={Asset selection based on high frequency Sharpe ratio},
	author={Wang, Christina Dan and Chen, Zhao and Lian, Yimin and Chen, Min},
	journal={Journal of Econometrics},
	volume={227},
	number={1},
	pages={168--188},
	year={2022},
	publisher={Elsevier}
}

@article{ying2022frechet,
	title={Fr{\'e}chet sufficient dimension reduction for random objects},
	author={Ying, Chao and Yu, Zhou},
	journal={Biometrika},
	volume={109},
	number={4},
	pages={975--992},
	year={2022},
	publisher={Oxford University Press}
}

@article{zhang2024dimension,
	title={Dimension reduction for Fr{\'e}chet regression},
	author={Zhang, Qi and Xue, Lingzhou and Li, Bing},
	journal={Journal of the American Statistical Association},
	volume={119},
	number={548},
	pages={2733--2747},
	year={2024},
	publisher={Taylor \& Francis}
}

@article{zhu2011model,
	title={Model-free feature screening for ultrahigh-dimensional data},
	author={Zhu, Li-Ping and Li, Lexin and Li, Runze and Zhu, Li-Xing},
	journal={Journal of the American Statistical Association},
	volume={106},
	number={496},
	pages={1464--1475},
	year={2011},
	publisher={Taylor \& Francis}
}

\end{document}